\documentclass[aps,prc,reprint,showpacs]{revtex4-1}

\usepackage{graphicx,color,rotating,pifont}
\usepackage{amsmath,amssymb,bm}
\usepackage{ae}
\usepackage{pstricks}

\DeclareMathSymbol{\NS}{\mathord}{AMSb}{"4E}
\DeclareMathOperator{\abrapar}{\big<}
\DeclareMathOperator{\aketpar}{\big>}

\newcommand{\ket}[1]{\ensuremath{\,|{#1}\rangle}}
\newcommand{\braket}[2]{\ensuremath{\langle{#1}|{#2}\rangle}}
\newcommand{\matrixe}[3]{\ensuremath{\langle{#1}|\,{#2}\,|{#3}\rangle}}
\newcommand{\rmatrixe}[3]{\ensuremath{ \abrapar {#1} \big|\big| \,{#2}\, \big|\big| {#3} \aketpar }}

\newcommand{\braketn}[1]{\ensuremath{\braket{#1}{#1}}}

\newcommand{\dmatrixe}[2]{\matrixe{#1}{#2}{#1}}

\newcommand{\expect}[1]{\ensuremath{\langle{#1}\rangle}}

\newcommand{\comm}[2]{\ensuremath{[{#1},{#2}]}}

\newcommand{\op}[1]{\ensuremath{#1}}
\newcommand{\adj}[1]{\ensuremath{{{#1}}^{\dag}}}

\renewcommand{\vec}[1]{\ensuremath{\bm{#1}}}

\newcommand{\sixj}[1]{
 \ensuremath{
   \begin{Bmatrix}
     #1
   \end{Bmatrix}
 }
}

\newcommand{\cO}{\ensuremath{\op{c}}}

\newcommand{\rO}{\ensuremath{\op{r}}}

\newcommand{\vO}{\ensuremath{\op{v}}}

\newcommand{\alphaO}{\ensuremath{\op{\alpha}}}

\newcommand{\ccO}{\ensuremath{\adj{\op{c}}}}

\newcommand{\aalphaO}{\ensuremath{\adj{\op{\alpha}}}}

\newcommand{\AO}{\ensuremath{\op{A}}}

\newcommand{\HO}{\ensuremath{\op{H}}}

\newcommand{\PO}{\ensuremath{\op{P}}}
\newcommand{\QO}{\ensuremath{\op{Q}}}
\newcommand{\RO}{\ensuremath{\op{R}}}
\newcommand{\TO}{\ensuremath{\op{T}}}

\newcommand{\VO}{\ensuremath{\op{V}}}

\newcommand{\OOO}{\ensuremath{\adj{\op{O}}}}

\newcommand{\pOV}{\ensuremath{\vec{\op{p}}}}

\newcommand{\rOV}{\ensuremath{\vec{\op{r}}}}

\newcommand{\ruOV}{\ensuremath{\hat{\vec{\op{r}}}}}

\newcommand{\DOV}{\ensuremath{\vec{\op{D}}}}

\newcommand{\ROV}{\ensuremath{\vec{\op{R}}}}

\newcommand{\jmu}{\ensuremath{j_\mu}}
\newcommand{\jjmu}{\ensuremath{j_{\mu'}}}
\newcommand{\jnu}{\ensuremath{j_\nu}}
\newcommand{\jjnu}{\ensuremath{j_{\nu'}}}

\newcommand{\lmu}{\ensuremath{l_\mu}}
\newcommand{\llmu}{\ensuremath{l_{\mu'}}}
\newcommand{\lnu}{\ensuremath{l_\nu}}
\newcommand{\llnu}{\ensuremath{l_{\nu'}}}

\newcommand{\hjmu}{\ensuremath{\widehat{j}_\mu}}
\newcommand{\hjjmu}{\ensuremath{\widehat{j}_{\mu'}}}
\newcommand{\hjnu}{\ensuremath{\widehat{j}_\nu}}
\newcommand{\hjjnu}{\ensuremath{\widehat{j}_{\nu'}}}

\newcommand{\hJ}{\widehat{J}}

\newcommand{\Tint}{\ensuremath{\TO_\text{int}}}

\newcommand{\Epair}{\ensuremath{E_\text{pair}}}

\newcommand{\Vlowk}{\ensuremath{V_{\text{low-k}}}}
\newcommand{\Vsrg}{\ensuremath{V_{\text{SRG}}}}

\newcommand{\aHO}{\ensuremath{a_{\text{HO}}}}

\newcommand{\half}{\ensuremath{\tfrac{1}{2}}}
\newcommand{\elem}[2]{\ensuremath{{}^{#2}\text{#1}}}
\newcommand{\nuc}[2]{\ensuremath{^{#2}\mathrm{#1}}}

\newcommand{\fm}{\ensuremath{\,\text{fm}}}

\newcommand{\keV}{\ensuremath{\,\text{keV}}}
\newcommand{\MeV}{\ensuremath{\,\text{MeV}}}

\newcommand{\GeV}{\ensuremath{\,\text{GeV}}}

\newcommand{\linemediumsolid}[1][black]{\unitlength1ex 
  {\color{#1}\begin{picture}(6,1)
  \linethickness{0.4mm}
  \put(0,0.5){\line(1,0){6.0}}
  \end{picture}}\nolinebreak
}

\newcommand{\linemediumdashed}[1][black]{\unitlength1ex 
  {\color{#1}\begin{picture}(6,1)
  \linethickness{0.4mm}
  \put(0,0.5){\line(1,0){1.5}}
  \put(2.2,0.5){\line(1,0){1.5}}
  \put(4.4,0.5){\line(1,0){1.5}}
  \end{picture}}\nolinebreak
}

\newcommand{\linemediumdotted}[1][black]{\unitlength1ex 
  {\color{#1}\begin{picture}(6,1)
  \linethickness{0.4mm}
  \put(0,0.5){\line(1,0){0.4}}
  \put(0.9,0.5){\line(1,0){0.4}}
  \put(1.8,0.5){\line(1,0){0.4}}
  \put(2.7,0.5){\line(1,0){0.4}}
  \put(3.6,0.5){\line(1,0){0.4}}
  \put(4.5,0.5){\line(1,0){0.4}}
  \put(5.4,0.5){\line(1,0){0.4}}
  \end{picture}}\nolinebreak
}

\newcommand{\symboldiamond}[1][black]{{\color{#1}$\blacklozenge$}}

\newcommand{\symbolbox}[1][black]{{\color{#1}$\blacksquare$}}
\newcommand{\symbolcircle}[1][black]{{\color{#1}\ding{108}}}

\newcommand{\symbolcross}[1][black]{{\color{#1}\ding{58}}}

\definecolor{FGViolet}{rgb}{0.61,0.32,0.61}
\definecolor{FGDarkBlue}{rgb}{0,0,0.6}
\definecolor{FGBlue}{rgb}{0,0,0.8}
\definecolor{FGLightBlue}{rgb}{0.2, 0.6, 0.8}
\definecolor{FGGreen}{rgb}{0.2,0.7,0.2}
\definecolor{FGLightGreen}{rgb}{0.4,1,0.4}
\definecolor{FGYellow}{rgb}{1,0.95,0}
\definecolor{FGOrange}{rgb}{0.95,0.5,0.1}
\definecolor{FGRed}{rgb}{0.8,0,0}
\definecolor{FGWhite}{rgb}{1,1,1}
\definecolor{FGLightGray}{rgb}{0.8,0.8,0.8}
\definecolor{FGGray}{rgb}{0.5,0.5,0.5}
\definecolor{FGDarkGray}{rgb}{0.3,0.3,0.3}
\definecolor{FGBlack}{rgb}{0,0,0}

\begin{document}
\title{Quasiparticle Random Phase Approximation with Interactions from the Similarity Renormalization Group}

\author{H. Hergert}
\email{hergert@nscl.msu.edu}
\affiliation{National Superconducting Cyclotron Laboratory, Michigan State University, East Lansing, MI 48824, USA}

\author{P. Papakonstantinou}
\email{Panagiota.Papakonstantinou@physik.tu-darmstadt.de}

\author{R. Roth}
\email{Robert.Roth@physik.tu-darmstadt.de}

\affiliation{Institut f\"ur Kernphysik, Technische Universit\"at Darmstadt,
64289 Darmstadt, Germany}

\date{\today}

\begin{abstract}
We have developed a fully consistent framework for calculations in the Quasiparticle Random Phase Approximation (QRPA) with $NN$ interactions from the Similarity Renormalization Group (SRG) and other unitary transformations of realistic interactions. The consistency of our calculations, which use the same Hamiltonian to determine the Hartree-Fock-Bogoliubov (HFB) ground states and the residual interaction for QRPA, guarantees an excellent decoupling of spurious strength, without the need for empirical corrections. While work is under way to include SRG-evolved $3N$ interactions, we presently account for some $3N$ effects by means of a linearly density-dependent interaction, whose strength is adjusted to reproduce the charge radii of closed-shell nuclei across the whole nuclear chart. As a first application, we perform a survey of the monopole, dipole, and quadrupole response of the calcium isotopic chain and of the underlying single-particle spectra, focusing on how their properties depend on the SRG parameter $\lambda$. Unrealistic spin-orbit splittings suggest that spin-orbit terms from the $3N$ interaction are called for. Nevertheless, our general findings are comparable to results from phenomenological QRPA calculations using Skyrme or Gogny energy density functionals. Potentially interesting phenomena related to low-lying strength warrant more systematic investigations in the future. 
\end{abstract}


\maketitle

\clearpage

\section{Introduction\label{sec:intro}}
In recent years, rare isotope beams have become a major focus of the experimental nuclear physics community. Using such beams, more and more exotic nuclei become accessible experimentally that exhibit novel structural features and excitations modes like neutron-skin vibrations and allow studies of sensitive details of the nuclear interactions and the theoretical models that are used to describe them. Over the past few decades, nuclear structure calculations for medium- and heavy-mass nuclei were almost exclusively carried out in the framework of Density Functional Theory (DFT), using phenomenological energy density functionals (EDFs) of the Skyrme or Gogny type \cite{Bender:2003bs}, or in Relativistic Mean-Field Theory \cite{Vretenar:2005ly}. While phenomenological EDFs uniformly describe the bulk properties of nuclei near the valley of stability very well, there is a significant model dependence and deterioration of quality for results in exotic nuclei and spectroscopic observables in general. While DFT formally resembles the Hartree-Fock (HF) and Hartree-Fock-Bogoliubov (HFB) methods, the EDF parametrization contain correlation effects beyond the mean field; on the one hand, this allows for a better description of experimental data in a comparatively simple framework, but on the other hand, there is no clear connection to the underlying $NN$ (and $3N, 4N$, \ldots) interactions, and, therefore, no way to improve the EDFs in a systematic fashion, e.g., by many-body perturbation theory (MBPT).

One way to overcome these problems is to use realistic $NN$ interactions like Argonne V18 \cite{Wiringa:1995or} or the potentials from next-to-next-to-next-to-leading order (N3LO) of chiral effective field theory (EFT) \cite{Epelbaum:2006mo,Entem:2003th}, which accurately describe $NN$ scattering data. The latter are particularly appealing because chiral EFT provides a consistent set of accompanying $3N$ interactions, although the $3N$ interaction has thus far only been derived to order N2LO \cite{Epelbaum:2006mo}. Since realistic interactions induce strong short-range correlations in the $NN$ system, one needs to tame their short-range behavior, preferably by means of a unitary transformation which automatically preserves the $NN$ observables. Examples of such unitary transformation techniques are the Unitary Correlation Operator Method (UCOM) \cite{Roth:2010vp}, and the Similarity Renormalization Group (SRG) approach \cite{Bogner:2010pq}, which will be the method explored in the following. The SRG evolution drives the two-body interaction to band-diagonality in momentum space, thereby decoupling low and high momenta. This decoupling results in soft interactions with greatly improved convergence properties in (quasi-)exact many-body methods. In addition, SRG-evolved interactions yield bound nuclei already at the mean-field level, and are suitable for low-order MBPT treatments (see Ref. \cite{Bogner:2010pq} and references therein).

Soft $NN$ interactions have been discussed in nuclear theory for many decades, but were originally discarded due to their inability to produce the proper saturation behavior in nuclear matter \cite{Bethe:1971qf}. From the modern point of view, this merely shows that for each $NN$ interaction, consistent $3N$ forces are required to properly describe nuclear systems. According to general EFT principles, up to $A$-body interactions must be considered in the $A$-nucleon system, for which chiral EFT guarantees a natural hierarchy with $NN > 3N > 4N$ etc.: the leading $3N$ force appears at N2LO, the leading $4N$ force at N3LO, and so on. The SRG evolution beautifully illustrates the inseparability of the nuclear interactions, because many-nucleon forces are naturally induced during the SRG flow \cite{Bogner:2010pq}. In the case of SRG-evolved $NN$ interactions, the inclusion of repulsive $3N$ interactions is essential to prevent overbinding in heavier nuclei. Formally, the $3N$ interaction must then be evolved consistently along with the $NN$ interaction, which has recently been accomplished by Jurgenson et al. \cite{Jurgenson:2009bs}. 

With realistic interactions and similarity transformation techniques, significant progress has been made toward a comprehensive description of nuclear structure all across the nuclear chart. \emph{Ab initio} calculations for light nuclei (see \cite{Roth:2010vp,Bogner:2010pq} and references therein) are complemented by mean-field based approaches in heavy nuclei. In recent years, we have developed a framework for using effective interactions, given in terms of their harmonic-oscillator matrix elements, in a wide-range of mean-field based approaches, from HF and HFB to the Random Phase Approximation (RPA) and its extensions \cite{Roth:2010vp,Paar:2006zf,Papakonstantinou:2007lu,Papakonstantinou:2010oq}. 

The purpose of this article is twofold. First, we extend the description of excitations to open-shell nuclei by means of a fully consistent Quasiparticle RPA \cite{Ring:1980bb,Suhonen:2007wo}: The QRPA is built on the ground states obtained from the HFB method, and the same intrinsic Hamiltonian, including the exact Coulomb interaction, is used in both the HFB and QRPA calculations. Second, we perform a survey of the response of isotopic chains using SRG-evolved $NN$ interactions. While work is under way to include SRG-evolved chiral $3N$ interactions in our overall framework, a $3N$ contact force (or equivalent density-dependent interaction) is implemented as an intermediate step, allowing us to carry out preparatory studies and identify issues in anticipation of the full $3N$ interaction. Our nuclear structure results based on SRG-evolved $NN$ (and eventually $3N$) interactions will provide important guidance for \emph{ab initio} DFT efforts in the framework of the Universal Nuclear Energy Density Functional (UNEDF) project \cite{web:UNEDF}.

This article is organized as follows. In Sec. \ref{sec:formalism} we review the QRPA formalism and provide details and tests of our implementation. In Sec. \ref{sec:ddi}, we fix the free parameter of the density-dependent interaction and discuss some open issues pertaining to $3N$ forces and proceed to summarize HFB results for the calcium isotopic chain in Sec. \ref{sec:hfb}. This sets the stage for the QRPA results, which are presented in Sec. \ref{sec:res}. Explicit expressions for the QRPA matrix elements are collected in the appendices.
\section{\label{sec:formalism}QRPA formalism and implementation}

\subsection{\label{sec:qrpa}Quasi-Particle Random Phase Approximation}
Our starting point is the intrinsic Hamiltonian
\begin{equation}
  \HO = \Tint + \VO\,,
\end{equation}
where the intrinsic kinetic energy is defined as \cite{Ring:1980bb,Hergert:2009wh}
\begin{equation}\label{eq:def_Tint}
   \Tint = \TO-\TO_\text{cm}=\left(1-\frac{1}{A}\right)\sum_i\frac{\pOV^2_i}{2m} - \frac{1}{mA}\sum_{i<j}\pOV_i\cdot\pOV_j\,.
\end{equation}
We formulate the QRPA in the canonical basis of the Hartree-Fock Bogoliubov (HFB) ground state \cite{Ring:1980bb,Paar:2003tr,Terasaki:2005ja}. Normal-ordering the Hamiltonian w.r.t. the HFB vacuum, we obtain
\begin{align}
   \HO &= E_0 + \sum_{kk'}H_{kk'}^{11} \aalphaO_k\alphaO_{k'}
	 + \VO_\text{res}
\end{align}
where $E_0$ is the energy expectation value in the HFB vacuum, and $\{\alphaO_k,\aalphaO_k\}$ are quasiparticle operators in the canonical basis. The residual interaction is given by
\begin{equation} \label{eq:def_Vres}
   \VO_\text{res} = \frac{1}{4}\sum_{kk'll'}\bar{v}_{kk'll'}:\ccO_k\ccO_{k'}\cO_{l'}\cO_l:\,,
\end{equation}
where $\ccO_k$ are the creation operators of the canonical basis in particle representation, and $\bar{v}_{kk'll'}$ denotes an antisymmetrized but not normalized two-body matrix element. Details on how to obtain the quasiparticle representation of $\VO_\text{res}$ can be found in Refs. \cite{Ring:1980bb,Suhonen:2007wo}.

Assuming spherical symmetry, the canonical basis states come in pairs $\{\ket{\mu, m_\mu}, \ket{\overline{\mu, m_\mu}}\}$ which are related by time reversal:
\begin{equation}
  \ket{\overline{\mu m}} = (-1)^{l+j-m}\ket{\mu -m}\,,
\end{equation}
where $\mu=(nlj\tau)$ is a collective index for the radial, angular momentum, and isospin quantum numbers. In the canonical basis, the Bogoliubov transformation between the particle and quasiparticle representation reduces to the BCS-like form \cite{Bogoliubov:1958pl,Valatin:1958wb}
\begin{subequations}\label{eq:def_qp}
  \begin{align}
    \aalphaO_{\mu m} &= u_{\mu} \ccO_{\mu m} + v_\mu \widetilde{\cO}_{\mu m}\,,\\
    \widetilde{\alphaO}_{\mu m} &= u_{\mu} \widetilde{\cO}_{\mu m} - v_\mu \ccO_{\mu m}\,,
  \end{align}
\end{subequations}
where we have expressed the annihilation operators as spherical tensors \cite{Edmonds:1957qt},
\begin{equation}\label{eq:def_sphadj}
   \widetilde{\alphaO}_{\mu m} = (-1)^{j+m}\alphaO_{\mu - m} = -(-1)^{l} \alphaO_{\overline{\mu m}}\,,
\end{equation}
and absorbed a factor $(-1)^{l}$ into the coefficients $v_\mu$ to simplify the formulas. 

The QRPA phonon creation operator in the canonical basis has the general form \cite{Rowe:1968eq}
\begin{equation}\label{eq:def_phonon_general}
   \OOO_k = \!\!\sum_{(\mu m) < (\mu' m')}\!\!X^k_{\mu m,\mu'm'}\aalphaO_{\mu m}\aalphaO_{\mu'm'} - Y^k_{\mu m,\mu'm'}\alphaO_{\mu'm'}\alphaO_{\mu m}\,,
\end{equation}
where the sum over quasiparticle states must be restricted to avoid double counting. Since we assume spherical symmetry, it is convenient to switch to an angular-momentum coupled representation \cite{Suhonen:2007wo}: \begin{equation}\label{eq:def_phonon_spherical}
   \OOO_{kJM} = \sum_{\mu \leq \mu'}X^{kJ}_{\mu\mu'}\mathcal{A}^\dag_{\mu\mu'JM} - Y^{kJ}_{\mu\mu'}\widetilde{\mathcal{A}}_{\mu\mu'JM}\,,
\end{equation}
where the coupled quasiparticle-pair creation operator is defined as
\begin{equation}
  \mathcal{A}^{\dag}_{\mu\mu'JM}\equiv\frac{1}{\sqrt{1+\delta_{\mu\mu'}}}\sum_{m,m'}
                     \braket{j m j'm'}{JM}\aalphaO_{\mu m}\aalphaO_{\mu'm'}
\end{equation}
and $\widetilde{\AO}_{\mu\mu'JM}$ is its spherical adjoint [cf. Eq. \eqref{eq:def_sphadj}].

Using the Equations-of-Motion method \cite{Rowe:1968eq,Suhonen:2007wo}, one can define the QRPA matrices $A$ and $B$ via the commutators ($\mu\leq\mu',\nu\leq\nu'$)
\begin{subequations}\label{eq:def_AB}
  \begin{align}
    A^{JM}_{\mu\mu',\nu\nu'}&\equiv\matrixe{\Psi}{\comm{\widetilde{\mathcal{A}}_{\mu\mu'JM}}{\comm{\HO}{\mathcal{A}^\dag_{\nu\nu'JM}}}}{\Psi}\,,
    \\
    B^{JM}_{\mu\mu',\nu\nu'}&\equiv\matrixe{\Psi}{\comm{\widetilde{\mathcal{A}}_{\mu\mu'JM}}{\comm{\HO}{\widetilde{\mathcal{A}}_{\nu\nu'JM}}}}{\Psi}\,,
  \end{align}
\end{subequations}
where we resort to the usual quasi-boson approximation by assuming that the many-body state $\ket{\Psi}$ is the HFB vacuum. For spherically symmetric systems, the QRPA matrices and the amplitudes $X$ and $Y$ are independent of the angular momentum projection, and one obtains the following reduced set of QRPA equations: 
\begin{equation}\label{eq:qrpa}
   \begin{pmatrix}
    A^J & B^J\\
   -B^{J*} & -A^{J*}
  \end{pmatrix}
  \begin{pmatrix}
   X^{kJ} \\ Y^{kJ}
  \end{pmatrix}
  = \hbar\omega_k
  \begin{pmatrix}
   X^{kJ} \\ Y^{kJ}
  \end{pmatrix}\,,
\end{equation}
where $\hbar\omega_k$ is the excitation energy of the $k$th QRPA state w.r.t. the ground state. Explicit expressions for the matrices $A$ and $B$ can be found in Appendix \ref{app:qrpa_mat}.

\subsection{Transition Operators\label{sec:qrpa_trans}}
For electric multipole transitions, the reduced transition probabilities are defined as
\begin{equation}\label{eq:def_bej}
   B(EJ, J_i\rightarrow J_f)\equiv\frac{1}{2J_i+1}\left|\rmatrixe{fJ_f}{\QO_J}{iJ_i}\right|^2\,.
\end{equation}
In the QRPA, we consider transitions from the $0^+$ ground state of an even-even nucleus to an excited state described by the QRPA phonon operator \eqref{eq:def_phonon_spherical}, and the reduced matrix element can be evaluated to (see, e.g., Ref. \cite{Suhonen:2007wo})
\begin{align}\label{eq:def_transme}
  &\rmatrixe{kJ}{\QO_J}{0}\notag\\
&=\sum_{\mu\leq\mu'}\frac{1}{\sqrt{1+\delta_{\mu\mu'}}}\left(u_\mu v_{\mu'}+(-1)^Jv_\mu u_{\mu'}\right)\notag\\
  &\quad\times
       \left(X_{\mu\mu'}^{kJ*}\rmatrixe{\mu}{\QO_J}{\mu'}+
                  (-1)^JY^{kJ*}_{\mu\mu'}\rmatrixe{\mu}{\QO_J}{\mu'}^*\right)\,.
\end{align}

In the limit of small momentum transfer, the multipole transition operator is defined as the sum of the isoscalar and isovector operators
\begin{equation}
  \QO^{IS}_{JM} = \tfrac{1}{2}e \sum_{i=1}^A \rO_i^J Y_{JM}(\ruOV_i)\,
\end{equation}
and
\begin{equation}  
  \QO^{IV}_{JM} = \tfrac{1}{2}e \sum_{i=1}^A \tau_3^{(i)}\rO_i^J Y_{JM}(\ruOV_i)\,.
\end{equation}
Exceptions are the monopole operator, which would be a constant unable to cause transitions, and therefore needs to be defined as
\begin{equation}
  \QO_{00} = e \sum_{i=1}^A \tfrac{1}{2}(1+\tau_3^{(i)})\rO_i^2 Y_{00}(\ruOV_i)\,,
\end{equation}
and the isoscalar and isovector dipole operators, which are corrected for center-of-mass effects \cite{Ring:1980bb}:
\begin{equation}\label{eq:E1_IS_mod}
  \QO^{IS}_{1M}=e\sum_{i=1}^A\left(\rO_i^3-\frac{5}{3}\expect{\RO_\text{ms}}\rO_i\right)Y_{1M}(\ruOV_i)\,,
\end{equation}
and
\begin{equation}\label{eq:E1_IV_mod}
  \QO^{IV}_{1M}=e\frac{N}{A}\sum_{p=1}^Z\rO_pY_{1M}(\ruOV_p)-e\frac{Z}{A}\sum_{n=1}^N\rO_nY_{1M}(\ruOV_n)\,,
\end{equation}
where $\RO_\text{ms}$ is the intrinsic mean-square radius operator \cite{Roth:2006lr}.

Since the $NN$ interactions used in this work are obtained by means of an SRG evolution, we need to address the issue of evolving observables in a consistent fashion. However, it has been demonstrated in related approaches like the UCOM \cite{Paar:2006zf} and the Lee-Suzuki transformation in the No-Core Shell Model \cite{Stetcu:2005qh} that the absolute values of the transition operator matrix elements entering \eqref{eq:def_bej} hardly change. The reason for this is the long-range, low-momentum character of the $\rO^J$ operator, while the SRG, UCOM, and Lee-Suzuki transformation modify the short-range, high-momentum matrix elements \cite{Anderson:2010br}. Since there are more significant uncertainties due to the $\lambda$-dependence of the interaction and the simple nature of the phenomenological $3N$ interaction, we content ourselves with using the unevolved transition operators in the following.

\subsection{Calculation Details}
Our QRPA implementation is suitable for use with NN Hamiltonians with or without a density-dependent NN (or contact 3N) term. In the present study we use an SRG-evolved Argonne V18 interaction supplemented with a phenomenological density-dependent two-body term to take into account missing genuine and induced three-nucleon interactions. 
The matrix elements of this potential, which we will denote as $\Vsrg$, and the two-body part of the intrinsic kinetic energy \eqref{eq:def_Tint} are evaluated in a relative spherical HO basis and transformed to the $jj$-coupled single-particle basis by using Talmi-Moshinsky brackets, as described in Ref. \cite{Roth:2005kk}. 
The matrix elements of the density-dependent interaction, which is described in Sec. \ref{sec:ddi}, can be evaluated directly in the $jj$-coupled basis (see, e.g., Ref. \cite{Hergert:2009zn} and also Appendix \ref{app:ddi}).

The input ground states for our QRPA calculations are obtained using the spherical HFB implementation described in Ref. \cite{Hergert:2009zn}: single-particle states are expanded in a spherical HO basis of 15 major oscillator shells, ensuring converged ground-state energies. In calcium isotopes, the typical ground-state energy gains from increasing the size of the single-particle basis from 11 to 13 major shells are about $200\,\keV$, and from 13 to 15 major shells $100\,\keV$ or less, corresponding to less than $0.1\%$ of the total energy. In the tin region, the absolute gains are roughly twice as large, but the relative accuracy is similar. The ground-state energies are minimized w.r.t. the oscillator length $\aHO$ by considering a mesh of values ranging from $1.5$ to $2.40\,\fm$ with a spacing of $0.05\,\fm$. For 15 oscillator shells, the ground-state energies of the adjacent $\aHO$ mesh points differ from those at the minima by $5\,\keV$ or less in the calcium chain, and  $40\,\keV$ or less in the tin chain.

\begin{figure}[t]
  \includegraphics[width=\columnwidth]{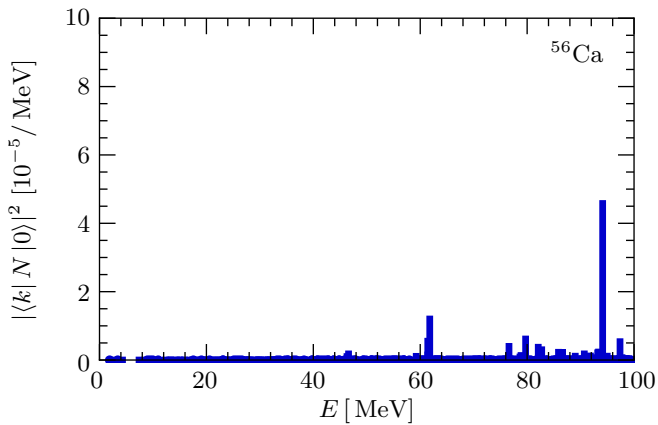}
  \caption{\label{fig:spur_numop}(Color online)
    Number operator response for nonspurious monopole states in $\nuc{Ca}{56}$ (see text).
    ($\Vsrg$+DDI with $\lambda=2.02\,\fm^{-1}, C_{3N}=3.87\,\GeV\fm^6$.)
  }
\end{figure}

After constructing the canonical basis of the HFB ground state, we determine all possible two-quasiparticle ($2qp$) configurations for a given $J^\pi$. We stress that the $2qp$ basis is not truncated in any way, which leads to an excellent decoupling of spurious states in our calculations (see Sec. \ref{sec:spurious}). To estimate uncertainties due to the discrete $\aHO$ mesh, we perform QRPA calculations for the HFB solutions at the neighboring points as well. The variation of both individual excited state and centroid energies is about $50\,\keV$ or less in the results presented in the following.

\subsection{Spurious States\label{sec:spurious}}
\begin{figure}[t]
  \includegraphics[width=\columnwidth]{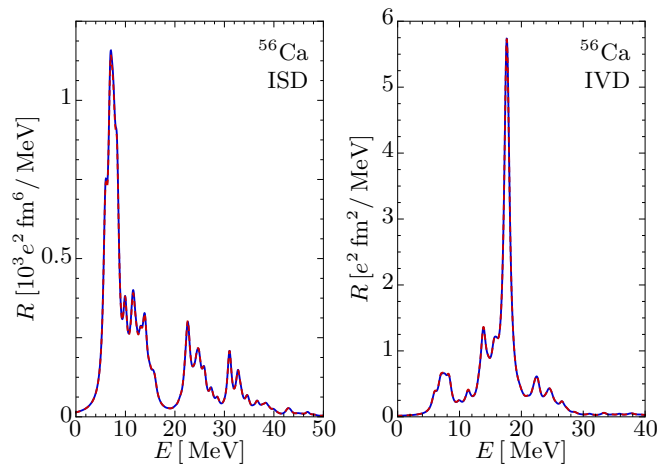}
  \caption{\label{fig:spur_dipole}(Color online)
    Isoscalar (left) and isovector (right) dipole strength distributions in $\nuc{Ca}{56}$ with (\linemediumsolid[FGBlue])\, and without (\linemediumdashed[FGRed])\, c.o.m. correction. The discrete strength distributions have been folded with a Lorentzian of width $\Gamma=1.0\MeV$. In the isoscalar channel, the spurious state has been removed explicitly from the response of the uncorrected operator (see text). ($\Vsrg$+DDI with $\lambda=2.02\,\fm^{-1}, C_{3N}=3.87\,\GeV\fm^6$.)
  }
\end{figure}

In a QRPA calculation, spurious states emerge as a consequence of the breaking of the symmetries of the nuclear Hamiltonian by the ground-state wavefunction. In the $0^+$ channel, nonvanishing neutron and/or proton pairing break the $U(1)$ symmetries associated with neutron and proton number (or, alternatively, nucleon number and charge) conservation. The use of the same interaction in the particle-hole and particle-particle channels in \emph{both} the HFB and QRPA calculation ensures that the corresponding spurious state(s) are well-decoupled from the excitation spectrum \cite{Paar:2003tr,Terasaki:2005ja}. 

As a typical example, we show the number operator response for the nonspurious $0^+$ states of $\nuc{Ca}{56}$ in Fig. \ref{fig:spur_numop}, which is less than $10^{-4}/\MeV$ in magnitude overall, and less than $10^{-6}/\MeV$ in the energy range up to 40 $\MeV$, which is relevant for the giant monopole resonance. The number operator response vanishes to machine accuracy if the pairing collapses (i.e., in the HF + RPA limit). The spurious $0^+$ state in $\nuc{Ca}{56}$ is found at $0.1\,\keV$, and the largest energies we found in our calculations are approximately $20\,\keV$, which still indicates excellent consistency. 

The decoupling is realized just as well for the spurious $1^-$ state associated with the breaking of translational invariance. To verify that our QRPA solutions are free of center-of-mass contamination, we compare the isoscalar $1^-$ strength distribution obtained with \eqref{eq:E1_IS_mod} and the uncorrected operator
\begin{equation}\label{eq:E1_IS}
   Q^{IS}_{1M}=e\sum_{i}^A r_{i}^3Y_{1M}(\ruOV_{i})\,.
\end{equation}
If translational invariance is properly restored for the QRPA solutions, the strength distributions must agree for the nonspurious states, so the only effect of the correction term in \eqref{eq:E1_IS_mod} is the removal of the spurious state associated with the translation of the whole nucleus.

In the isovector case, the corrected dipole operator \eqref{eq:E1_IV_mod} is used, which is equivalent to 
\begin{equation}\label{eq:E1_mod}
  \DOV' = \sum_{p}^Z e \left(\rOV_p - \ROV \right) = \frac{N}{A}e\sum_{p}^{Z} \rOV_p - \frac{Z}{A}e\sum_{n}^{N}\rOV_n\,.
\end{equation}
If translational symmetry is properly restored, the c.o.m. operator $\ROV$ cannot cause transitions, and thus for intrinsic excitations the matrix elements of $\DOV'$ must be identical to those of
\begin{equation}\label{eq:E1}
  \DOV = \sum_{p}^{Z} e \rOV_p\,.
\end{equation}
Figure \ref{fig:spur_dipole} demonstrates that isoscalar and isovector strength distributions of the corrected operators are practically identical to those of the operators \eqref{eq:E1_IS} and \eqref{eq:E1}, respectively. This confirms our previous findings for closed-shell nuclei \cite{Paar:2006zf} in the more general HFB+QRPA framework for open-shell nuclei.

As in the monopole case, the spurious $1^-$ state itself lies at very low energies, independent of the mass. It is found between $1$ and $5\,\keV$ in closed-shell nuclei, and below $20\,\keV$ for open-shell nuclei in all cases. These energies are considerably lower than the spurious $1^-$ energies of several hundred $\keV$ which are reported for other consistent QRPA approaches in the literature \cite{Terasaki:2005ja,Peru:2005ys,Peru:2008vn,Losa:2010kx}. The reason is the use of the intrinsic kinetic energy \eqref{eq:def_Tint} in our HFB+QRPA calculations. If we do not subtract $\TO_\text{cm}$, the spurious state energies increase to the sizes reported by other groups, but the quality of the translational-symmetry restoration is not affected.   
 
\section{\label{sec:ddi}The Density-Dependent Interaction}

In the present study, we supplement an SRG-evolved Argonne V18 $NN$ interaction by a phenomenological density-dependent two-body term to account for missing genuine and induced $3N$ interactions. As discussed in Ref. \cite{Waroquier:1976go}, the linearly density-dependent interaction (DDI)
\begin{equation}\label{eq:def_ddi}
   \vO[\rho] = \frac{C_{3N}}{6}\left(1+\PO_\sigma\right)\rho\left(\frac{\rOV_1+\rOV_2}{2}\right)\delta^3\left(\rOV_1-\rOV_2\right)\,,
\end{equation}
where $\PO_\sigma$ is the spin-exchange operator, gives the same contribution to the ground-state energy as the $3N$ contact interaction
\begin{equation}\label{eq:def_3Ncontact}
   \vO_3 = C_{3N}\delta^3\left(\rOV_1-\rOV_2\right)\delta^3\left(\rOV_2-\rOV_3\right)
\end{equation}
in systems with time-reversal invariance. Such a contact force was recently used in conjunction with similarity-transformed interactions in Hartree-Fock and MBPT calculations \cite{Guenther:2010ge}. 

Since the QRPA is the limit of low-amplitude motion of time-dependent HFB, the QRPA matrix in Eq. \eqref{eq:qrpa} is (up to the metric) the stability matrix of the ground-state energy functional \cite{Ring:1980bb}, and the replacement of the $3N$ contact term with the DDI \eqref{eq:def_ddi} is meaningful in this context, provided one properly takes rearrangement terms due to the density-dependence into account (see Ref. \cite{Waroquier:1987ce} and Appendix \ref{app:ddi}).

In the present work, we treat the strength of the density-dependent term as a running coupling constant which depends on the SRG parameter $\lambda$, since it is an effective parametrization of initial $3N$ interactions, which should be present in the ``bare'' nuclear Hamiltonian, as well as induced $3N$ interactions, which are generated during the SRG flow (see Ref. \cite{Bogner:2010pq} and references therein). We fix $C_{3N}(\lambda)$ in Hartree-Fock calculations by fitting the experimental charge radii of a set of closed-shell nuclei, because the radii are much less sensitive to many-body corrections than the ground-state energy.  

\begin{table}[t]
   \begin{tabular}{c@{\extracolsep{10pt}}c@{\extracolsep{10pt}}c}
      \hline\hline
      &$\lambda [\fm^{-1}]$ & $C_{3N}(\lambda) [\GeV\fm^6]$ \\
      \hline
      &1.78 & 4.41  \\
     $\mathrm{Ca}$ & 2.02 & 3.87  \\
      & 2.40 & 2.94 \\
      \hline
      &1.78 & 4.95 \\
     $\mathrm{Sn}$ & 2.02 & 4.35 \\
      & 2.40 & 3.42  \\
      \hline\hline
   \end{tabular}
   \caption{\label{tab:lambda_C3N}
     Running coupling strength $C_{3N}(\lambda)$ for various values $\lambda$ used in this work (see text).
   }
\end{table}

\begin{figure}[t]
  \includegraphics[width=\columnwidth]{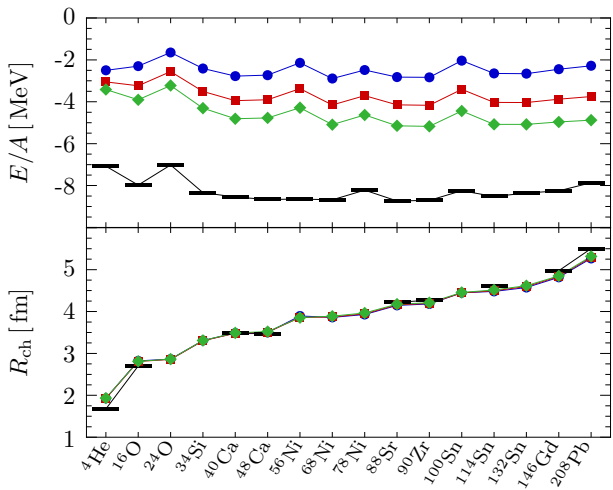}\\[-30pt]
  \caption{\label{fig:srgXXXX_ddXXX_stdnucl}(Color online)
    Ground-state energy per nucleon and charge radii of closed shell nuclei for $\Vsrg$+DDI with $(\lambda [\fm^{-1}], C_{3N} [\GeV\fm^6])$= (2.40, 2.94) (\symbolcircle[FGBlue]), (2.02, 3.87) (\symbolbox[FGRed]), and (1.78, 4.41) (\symboldiamond[FGGreen]). Experimental values \cite{Audi:2002af,Angeli:2004ts} are indicated by black bars.
  }
\end{figure}

In Table \ref{tab:lambda_C3N}, we list the $C_{3N}(\lambda)$ values for three $\lambda$'s used in the following; Fig. \ref{fig:srgXXXX_ddXXX_stdnucl} shows the corresponding ground-state energies (per nucleon) and charge radii for the fit nuclei. We find that for these nuclei, the $\lambda$-dependence of the charge radii can be absorbed into $C_{3N}(\lambda)$ at the HF(B) level, while a many-body approach beyond the mean-field and a complete treatment of the $3N$ interaction is required to reduce or remove the $\lambda$-dependence of the ground-state energies. 

The strength of our DDI is notably lower than the strength of the density-dependent term of past and current Skyrme functionals (see Ref. \cite{Kortelainen:2010vm} and references therein). In part, such a strong repulsive DD term is required to counteract the attractive terms in the Skyrme functional that are necessary to fit binding energies and radii at the same time in a mean-field calculation, which differs from our fit strategy.  

We also note that we do not need to use fractional density dependencies for the DDI to obtain reasonable GMR energies in Sec. \ref{sec:qrpa}. Modern Skyrme functionals use fractional density dependencies because the strong linear density-dependent terms tend to overestimate the GMR energy, i.e., the incompressibility of nuclear matter. Since phenomenological Skyrme functionals with constant coefficients correspond to nuclear contact interactions, non-integer density dependencies parametrize physical effects from finite-range $NN$ interactions in addition to the 3N contact term \cite{Stoitsov:2010vn}. 

Inspecting Fig. \ref{fig:srgXXXX_ddXXX_stdnucl}, we note that the quality of the fit to the charge radii varies with $A$: the charge radius of $\nuc{O}{16}$ is slightly too large, while the charge radii of heavier nuclei are underestimated. This is evidence that the limited spin-isospin dependence of the DDI \eqref{eq:def_ddi}, which only acts in the $(S,T)=(1,0)$ channel, is insufficient. Explicit spin and isospin degrees of freedom in the $3N$ interaction --- as in the chiral $3N$ interaction, for instance --- would lead to non-vanishing DDI matrix elements in the other $(S,T)$ channels \cite{Holt:2010dn}, and allow for a different scaling with the particle number. Since we will be focusing on the calcium isotopic chain and selected tin isotopes in the remainder of this work, we list two sets of DDI parameters in Table \ref{tab:lambda_C3N} that were optimized for the corresponding mass regions by fitting the charge radii of $\nuc{Ca}{40}$ and $\nuc{Sn}{114}$, respectively.

\begin{table}[t]
\begin{tabular*}{\columnwidth}{c@{\extracolsep\fill}c  c c c c  c}\hline\hline
        & Level & \multicolumn{4}{c}{\rule[-1mm]{0mm}{5mm}$(\lambda [\fm^{-1]},C_{3N} [\GeV\fm^6])$} & Exp. \\
        &         & (2.4, -) & (2.4, 2.94) & (2.02, 3.87) & (1.78, 4.41) & \\
\hline

\elem{O}{16}   & $\pi\,0p$ & \phantom{0}9.61   &  3.15  &  3.12 &  3.02 &  6.32 \\
               & $\nu\,0p$ & 10.04   &  3.18  &  3.14 &  3.05 &  6.18 \\
\hline
\elem{Ca}{40}  & $\pi\,0d$ & 15.19   &  4.21  &  4.14 &  4.04 &  6.00 \\
               & $\pi\,0f$ & 12.07   &  3.78  &  4.10 &  4.18 &  4.95 \\
               & $\nu\,0d$ & 15.95   &  4.27  &  4.19 &  4.08 &  6.00 \\
               & $\nu\,0f$ & 14.90   &  4.39  &  4.54 &  4.53 &  4.88 \\
\hline
\elem{Ca}{48}  & $\nu\,0f$ & 15.96   &  3.51  &  3.46 &  3.36 &  8.97 \\  
\hline
\elem{Sn}{100} & $\pi\,0g$ & 16.46   &  2.45  &  2.43 &  2.39 &  6.82 \\  
               & $\nu\,0g$ & 17.63   &  2.03  &  1.99 &  1.96 &  7.00 \\  
\hline
\elem{Sn}{132} & $\pi\,0g$ & 15.61   &  1.82  &  1.71 &  1.67 &  6.08 \\  
               & $\nu\,0h$ & 20.12   &  2.98  &  2.89 &  2.48 &  6.53 \\
\hline\hline
\end{tabular*}
\caption{\label{tab:spinorbit}
Proton ($\pi$) and neutron ($\nu$) spin-orbit splittings in $\MeV$ for different interaction parameters $(\lambda, C_{3N})$, compared to experimental values \cite{Isakov:2002bn,Isakov:2004fk}.}
\end{table}

A serious issue emerges for the spin-orbit splittings, which are listed for various nuclei in Table \ref{tab:spinorbit}. Using just a two-body SRG-evolved $NN$ interaction with $\lambda=2.40\,\fm^{-1}$, the spin-orbit splittings are about 2--3 times as large as values extracted from experiment. HF calculations for $\Vsrg$ with $\lambda=1.78$ and $2.02\,\fm^{-1}$ collapse and are not included. The repulsive DDI stabilizes the HF(B) calculations and leads to a compression of the single-particle spectra, but it results in a significant underestimation of the spin-orbit splittings. As a consequence, there is a shift in the major shell closures:  $\nuc{Ca}{48}$, for instance, has a small non-vanishing neutron pairing energy because the $0f_{7/2}$ level lies  close to the $pf$ major shell, while $\nuc{Sn}{120}$ is essentially a closed-shell nucleus because the $0h_{11/2}$ level is not shifted to sufficiently low energies to produce the $N=82$ major shell closure. 

The resulting proximity of levels with $\Delta j=\Delta l=2$ in the region of the Fermi surface favors strong quadrupole interactions, and causes the nucleus to develop a deformation. In our spherical QRPA calculations, this instability w.r.t. to quadrupole deformations is signaled by the lowest $2^+$ energy becoming purely imaginary, and we have to discard results based on the unstable spherical ground-state configuration. The calcium isotopic chain is free of this pathology, hence we focus on these isotopes in the following, and defer calculations for heavier isotopic chains to a future publication. Since the present work sets the stage for the use of the chiral $3N$ interaction (or a density-dependent variant), which contains additional spin-orbit and tensor terms (see, e.g., Refs. \cite{Kaiser:2008oc,Kaiser:2010dz,Holt:2010dn}), we expect a significant impact on the spin-orbit physics, but it remains to be seen whether the discussed problem can be resolved.

\section{Ground-State Properties\label{sec:hfb}}

\begin{figure*}[t]
  \includegraphics[width=2\columnwidth]{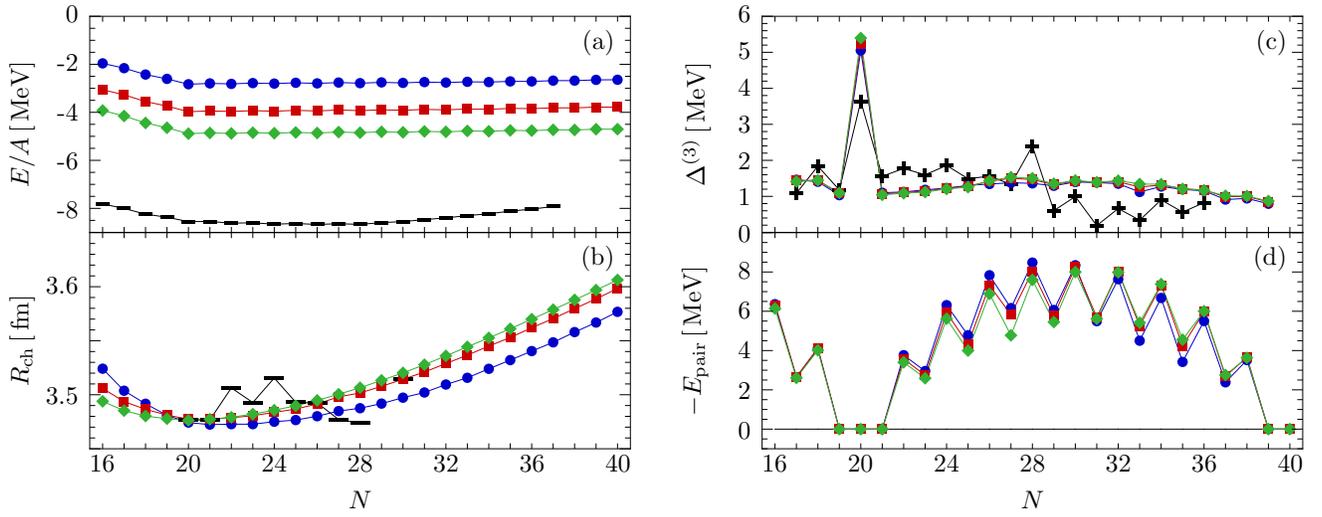}
  \caption{\label{fig:srgXXXX_ddXXX_CaXX}(Color online) Ground-state properties of the calcium isotopes
    for $\Vsrg$+DDI with $(\lambda [\fm^{-1}], C_{3N} [\GeV\fm^6])$= (2.40, 2.94) (\symbolcircle[FGBlue]), (2.02, 3.87) (\symbolbox[FGRed]), and (1.78, 4.41) (\symboldiamond[FGGreen]): (a) ground-state energies per nucleon, (b) charge radii, (c) odd-even mass differences, and (d) pairing energies. Experimental values \cite{Audi:2002af,Angeli:2004ts} are indicated by black bars or crosses.
  }
\end{figure*}

Before we present results from our QRPA calculations, we briefly discuss the properties of the calcium ground states on which the QRPA is built. All results have been obtained with SRG-evolved Argonne V18 interactions, supplemented by the adjusted DDI discussed in Sec. \ref{sec:ddi}. Odd nuclei have been treated in a self-consistent Equal Filling Approximation (EFA) \cite{Perez-Martin:2008zf}.

In Fig. \ref{fig:srgXXXX_ddXXX_CaXX}, we summarize the HFB ground-state properties of the studied calcium isotopes. While the elimination of short-range correlations via the SRG evolution yields bound nuclei for $\Vsrg$+DDI at the HFB level, ground-state energies shown in Fig. \ref{fig:srgXXXX_ddXXX_CaXX}a are underestimated by 4-6 $\MeV$ per nucleon, depending on the chosen $\lambda$. As discussed, e.g., in Ref. \cite{Roth:2010vp}, this is due to long-range correlations that are not taken into account by HFB, but can be recovered to a large extent by low-order MBPT due to the perturbative character of the SRG-evolved $NN$ interaction \cite{Bogner:2010pq}. 

Figure \ref{fig:srgXXXX_ddXXX_CaXX}b displays the charge radii, which are reasonably close to experimental values due to the fit of the DDI discussed in Sec. \ref{sec:ddi}. Overall, our calculations fail to reproduce the pattern of the experimental charge radii between $\nuc{Ca}{40}$ and $\nuc{Ca}{48}$, whose proper description requires the inclusion of effects beyond the mean-field (see, e.g., Ref. \cite{Bhattacharya:1993ij} and references therein). For $\nuc{Ca}{40}$ and $\nuc{Ca}{48}$, though, they are not far off.

In Figs. \ref{fig:srgXXXX_ddXXX_CaXX}c and \ref{fig:srgXXXX_ddXXX_CaXX}d, we show the experimental odd-even binding energy differences
\begin{equation}
   \Delta^{(3)} = \frac{(-1)^N}{2}\left(E(N+1)-2E(N)+E(N-1)\right)
\end{equation}
and the pairing energies (expressed in the canonical basis) \cite{Ring:1980bb}
\begin{equation}
   \Epair = \frac{1}{2}\sum_\mu (2j_\mu+1)\Delta_{\mu\mu}u_\mu v_\mu\,,
\end{equation}
respectively. $\Delta^{(3)}$, in particular, is sensitive to the single-particle structure of the HFB vacuum state near the Fermi surface. 

\begin{figure}[b]
  \includegraphics[width=\columnwidth]{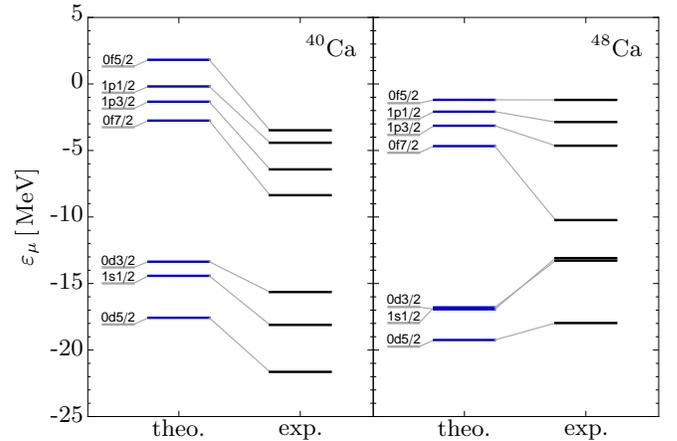}
  \caption{\label{fig:srg0600_dd645_Ca_Ecann}(Color online)
    Canonical neutron single-particle energies in $\nuc{Ca}{40}$ and $\nuc{Ca}{48}$ for $\Vsrg$+DDI with $\lambda=2.02\,\fm^{-1},C_{3N}=3.87\GeV\fm^6$. Experimental values have been taken from Refs. \cite{Isakov:2002bn,Isakov:2004fk}.
  }
\end{figure}

\begin{figure*}[t]
  \includegraphics[width=2\columnwidth]{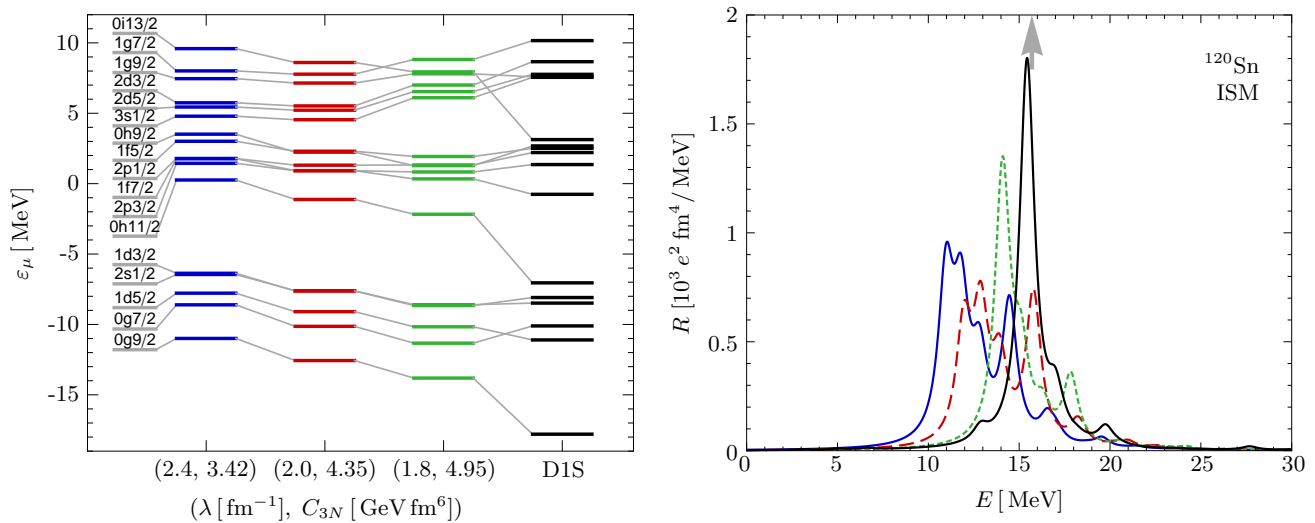}
  \caption{\label{fig:srgXXXX_ddXXX_Sn120}(Color online)
    Canonical single-neutron energies (left) and isoscalar monopole response (right) of $\nuc{Sn}{120}$ for $\Vsrg$+DDI with $(\lambda [\fm^{-1}], C_{3N} [\GeV\fm^6])$= (2.40, 3.42) (\linemediumsolid[FGBlue]), (2.02, 4.35) (\linemediumdashed[FGRed]), and (1.78, 4.95) (\linemediumdotted[FGGreen]). Results obtained with Gogny D1S are included for comparison (\linemediumsolid), and the arrow indicates the experimental centroid energy \cite{Li:2010jj}. The discrete ISM strength distributions have been folded with a Lorentzian of width $\Gamma=1\,\MeV$.
  }
\end{figure*}

We find that the agreement with experimental values is reasonable in the range $16\leq N < 28$, although the staggering is less pronounced in our calculation, which is at least in part due to the lack of time-reversal symmetry breaking in our EFA treatment of odd nuclei. The sharp jump in the $\Delta^{(3)}$ at $N=20$ is a signal of the major shell closure, which leads to the expected collapse of HFB pairing in $\nuc{Ca}{40}$. A similar jump is expected at the $N=28$ shell closure, but here the experimental data are not reproduced, and we note that the pairing energy does not vanish in $\nuc{Ca}{48}$. A look at the canonical single-neutron energies shown in Fig. \ref{fig:srg0600_dd645_Ca_Ecann} helps to clarify this issue. In our calculation, the spin-orbit splitting of the $0f$ levels in $\nuc{Ca}{48}$ is about half the experimental splitting. Consequently, the $0f_{7/2}$ level lies close to the main $pf$-shell, and the density of unoccupied levels is sufficiently high to produce pairing. We also note that the underestimation of the spin-orbit splittings is strongly isospin dependent, because the splittings in he $N=Z$ nucleus $\nuc{Ca}{40}$ are reproduced fairly well by comparison (also cf. Table \ref{tab:spinorbit}). As indicated in Sec. \ref{sec:ddi}, the issue of missing spin-orbit strength is not limited to the calcium isotopic chain or specific sets of interaction parameters $\lambda$ and $C_{3N}$, and it becomes an obstacle for QRPA calculations in heavier nuclei.

We conclude this section by mentioning that the present results for the odd-even binding energy differences are compatible with those of ``hybrid'' studies where $\Vsrg$ or $\Vlowk$ are used as pairing interactions in conjunction with phenomenological energy density functionals like SLy4 \cite{Lesinski:2009ty} or Gogny D1S \cite{Hergert:2009zn}. The inclusion of the DDI overcomes the low level density that obstructs pairing in HFB calculations when only similarity-transformed two-body interactions are used \cite{Hergert:2009zn}, and brings the single-particle energies closer to EDF results. Since the DDI \eqref{eq:def_ddi} vanishes in the pairing channel by construction, the $\Delta^{(3)}$ and pairing energies shown in Figs. \ref{fig:srgXXXX_ddXXX_CaXX}c and \ref{fig:srgXXXX_ddXXX_CaXX}d probe only the pairing properties of $\Vsrg$, and we obtain magnitudes that are similar to the hybrid approaches, aside from issues with the systematics due to the missing spin-orbit strength. It will be interesting to see how the situation changes once a consistent set of SRG-evolved $NN$ and $3N$ interactions is used in the calculation, because the latter will directly affect the pairing channel.

\section{Nuclear Response\label{sec:res}}

\subsection{General effects of the SRG Evolution\label{sec:res_srg}}
We first discuss the general effects of the SRG evolution on the nuclear response, using the isoscalar $0^+$ response of $\nuc{Sn}{120}$ as an example. $\nuc{Sn}{120}$ is chosen because it has a richer single-particle structure around the Fermi surface than the calcium isotopes, which facilitates the discussion, and because its ground state is stable against quadrupole deformation in our calculations (cf. Sec. \ref{sec:ddi}).

\begin{figure*}[t]
  \includegraphics[width=2\columnwidth]{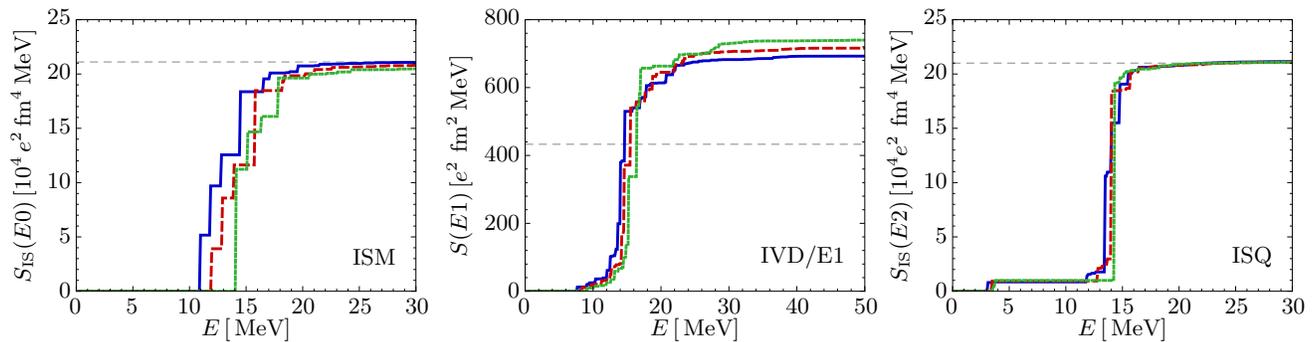}
  \caption{\label{fig:srgXXXX_ddXXX_Sn120_sum}(Color online)
    Running energy-weighted sums of $\nuc{Sn}{120}$ for $\Vsrg$+DDI with $(\lambda [\fm^{-1}], C_{3N} [\GeV\fm^6])$= (2.40, 3.42) (\linemediumsolid[FGBlue]), (2.02, 4.35) (\linemediumdashed[FGRed]), and (1.78, 4.95) (\linemediumdotted[FGGreen]). The light gray dashed lines indicate the values of the classical sum rules (see text). 
  }
\end{figure*}

In Fig. \ref{fig:srgXXXX_ddXXX_Sn120}, we show the canonical single-neutron energies and the isoscalar monopole response function for $\Vsrg$+DDI with the three parameter sets that have been optimized for the tin isotopic chain (cf. Table \ref{tab:lambda_C3N}). We note that the shell structure of the single-neutron energies becomes more pronounced as $\lambda$ is lowered. There are two distinct effects: During the SRG evolution, sub-shells move closer together, while the gaps between major shells increase as the SRG evolution renders the interaction increasingly nonlocal and thereby reduces the effective mass \cite{Roth:2006lr,Hergert:2009zn,Roth:2010vp}. For comparison, we have also calculated the single-neutron spectrum of $\nuc{Sn}{120}$ using the Gogny D1S functional. Except for the underestimation of the $0g$, $0h$, and $0i$ spin-orbit splittings, the single-neutron spectra of the three $\Vsrg$+DDI and D1S are rather similar for the levels around the Fermi surface. The similar level density for $\Vsrg$+DDI and for Gogny D1S, which corresponds to an effective mass of about $0.7$ times the bare nucleon mass, suggests that the inclusion of the repulsive DDI compensates for the extremely low effective mass that is a common feature of low-momentum interactions.

\begin{table}[b]
   \begin{tabular}{c@{\extracolsep{10pt}}c@{\extracolsep{10pt}}c@{\extracolsep{10pt}}c}
      \hline\hline
      & & \multicolumn{2}{c}{$m_{1}/m_{0}\,[\MeV]$}\\
      $\lambda [\fm^{-1}]$ & $C_{3N}(\lambda) [\GeV\fm^6]$ & HF & QRPA \\
      \hline
      1.78 & 4.95 & 23.5 & 15.2 \\
      2.02 & 4.35 & 22.5 & 14.2 \\
      2.40 & 3.42 & 21.3 & 12.9 \\
      \hline\hline
   \end{tabular}
   \caption{\label{tab:Sn120_centroids}
     Centroid energies of the isoscalar monopole strength distribution in $\nuc{Sn}{120}$ for different $\Vsrg$+DDI. 
     The energy integration intervals for $m_{0/1}$ were $5-50\,\MeV$ for the unperturbed HF response and $5-25\,\MeV$ for QRPA (see text).
   }
\end{table}

The isoscalar monopole (ISM) response functions displayed in Fig. \ref{fig:srgXXXX_ddXXX_Sn120} reflect the evolution of the single-particle spectra with the $\lambda$. The significant fragmentation of the ISM strength distribution for $\lambda=2.40\,\fm^{-1}$ is reduced notably due to the bunching of levels within the major shells as $\lambda$ is lowered. At the same time, the gap between major shells increases, leading to a shift of the strength distribution to higher energies. This shift to higher energies competes with the attractive isoscalar residual interaction, which aims to shift the response to lower energies. To get some insight into how the two effects depend on $\lambda$, we list the centroid energies for the unperturbed HF response without the residual interaction, and the QRPA response in Table \ref{tab:Sn120_centroids}. The integration intervals for the moments $m_{0/1}$ have been chosen by inspecting plots of the response functions. In the HF case (not shown), the bulk of the ISM strength lies between 10 and $30\,\MeV$, with a tail containing a few percent of the energy-weighted sum rule extending up to roughly $45\,\MeV$. In the full QRPA calculation, the ISM strength is concentrated in a much smaller range from $10$ to $\sim25\,\MeV$ due to the attractive residual interaction (see Fig. \ref{fig:srgXXXX_ddXXX_Sn120}), and the centroid lies $8.3$ to $8.4\,\MeV$ below the centroid of the unperturbed HF response for the three considered parameter sets. The uniformity of this difference in the centroid energies suggests that there is only little change in the isoscalar residual interaction in the studied range of $\lambda$'s, and the positive shift due to the increase of the major shell gap is the dominant effect. 

The ISM response for $\Vsrg$+DDI with $\lambda=1.78\,\fm^{-1}$ and Gogny D1S is rather similar. This can be seen as confirmation of the effective low-momentum nature of the Gogny functionals. Furthermore, the similarity of the response also shows that there is a decoupling between the static and collective dynamical properties of the nucleus, because the HFB ground-state energies obtained with Gogny D1S are very close to experimental ground-state energies in the tin isotopes, whereas $\Vsrg$+DDI underestimates the ground-state energies by several $\MeV$ per nucleon, strongly depending on the SRG parameter $\lambda$. The ground-state energy is an absolute quantity, while excitations are relative quantities that primarily depend on energy differences, and consequently, a simultaneous global shift of the levels involved in any single-particle transition will not alter the excitation energy. Such a shift seems to account for the bulk of the ground-state energy differences between $\Vsrg$+DDI and Gogny D1S. 

The present findings are in concordance with previous RPA studies of closed-shell nuclei with UCOM interactions. In Ref. \cite{Papakonstantinou:2007lu}, in particular, the RPA was formulated using the exact RPA ground state, and it was shown that the RPA ground-state correlations have little impact on the response. The RPA ground-state energy, however, contains many-body corrections \cite{Ellis:1970fv}, including second-order MBPT diagrams that recover much of the missing ground-state energy due to the perturbative nature of SRG and UCOM interactions \cite{Roth:2010vp,Guenther:2010ge}. 
These observations are further indications that higher-order many-body corrections to the ground state are indeed small if SRG or UCOM interactions are used, but a more detailed and quantitative study of this subject would be of great interest.

\begin{figure*}[t]
  \includegraphics[width=2\columnwidth]{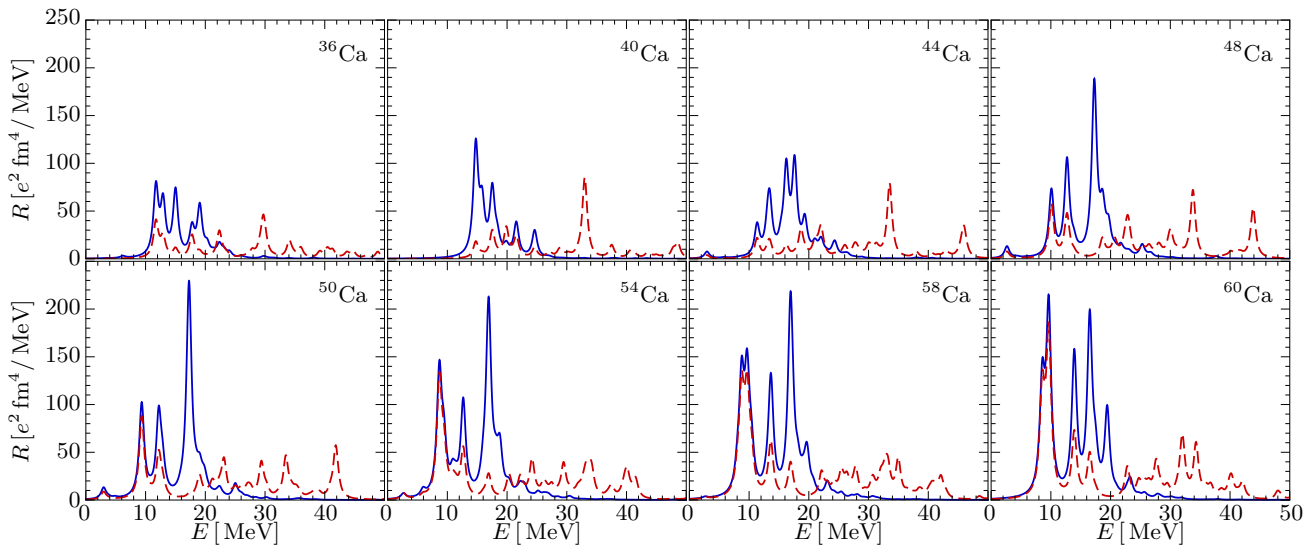}
  \caption{\label{fig:srg0600_dd645_CaXX_ISVM}(Color online)
    Isoscalar (\linemediumsolid[FGBlue]) and isovector monopole response (\linemediumdashed[FGRed]) of selected calcium isotopes for $\Vsrg$+DDI with $(\lambda [\fm^{-1}], C_{3N} [\GeV\fm^6])$= (2.02, 3.87). The discrete strength distributions have been folded with a Lorentzian of width $\Gamma=1\,\MeV$.
  }
\end{figure*}

We conclude this section by investigating the effect of the SRG evolution on the energy-weighted sum rules (EWSRs). For the isoscalar monopole and quadrupole channels, there are the well-known classical expressions (see, e.g., Ref. \cite{Ring:1980bb})
\begin{align}\label{eq:ism_sum}
  S_\text{IS}(E0)&=\frac{2\hbar^2e^2}{m}\left(N\expect{\rO^{2}_{n}}+Z\expect{\rO^{2}_{p}}\right)\,,\\
  S_\text{IS}(E2)&=\frac{25\hbar^2e^2}{4\pi m}\left(N\expect{\rO_n^{2}}+Z\expect{\rO_p^{2}}\right)\,,\label{eq:isq_sum}
\end{align}
where $\expect{\rO^{2}_{n/p}}$ are the intrinsic neutron and proton mean-square radii of the HFB solution, and for the dipole strength, we have the Thomas-Reiche-Kuhn sum rule \cite{Ring:1980bb}
\begin{equation}\label{eq:trk_sum}
  S(E1)=\frac{\hbar^2e^2}{2m}\frac{9}{4\pi}\frac{NZ}{A}\,.
\end{equation}
Equations \eqref{eq:ism_sum} to \eqref{eq:trk_sum} are derived by assuming that the interaction commutes with the transition operators, which is only the case for local interactions without isospin exchange contributions. Thus, the deviation of the calculated energy-weighted sum from the classical expressions is a measure for the size of momentum-dependent and isospin exchange contributions in the employed $\Vsrg$+DDI.

In Fig. \ref{fig:srgXXXX_ddXXX_Sn120_sum}, we show the running energy-weighted sum in the isoscalar monopole and quadrupole (ISQ), and the isovector dipole (IVD) channels. As discussed in Sec. \ref{sec:qrpa_trans}, the latter also corresponds to the running energy-weighted E1 strength at low momentum transfer. The running sums exhibit general features that reflect the findings for the ISM response function shown in Fig. \ref{fig:srgXXXX_ddXXX_Sn120}, i.e., as $\lambda$ is lowered, the strength becomes less fragmented, causing fewer jumps in Fig. \ref{fig:srgXXXX_ddXXX_Sn120_sum}, and the running starts at a higher excitation energy. 

At excitation energies of $\sim 30\,\MeV$, the isoscalar sum rules saturate, and Eqs. \eqref{eq:ism_sum} and \eqref{eq:isq_sum} are almost completely exhausted. In general, the quadrupole response exhibits a much weaker $\lambda$-dependence than the monopole and dipole response, which is presumably due to the surface excitation character of the former, while the latter are volume modes.

The total IVD sum exhibits the most pronounced $\lambda$-dependence of the three investigated cases. For $\lambda=2.40\,\fm^{-1}$, the dipole EWSR is enhanced by 60\% over the TRK value obtained from Eq. \eqref{eq:trk_sum}, and since strength is shifted to momentum-dependent terms as $\lambda$ is lowered, the enhancement increases to 70\% for $\lambda=1.78\,\fm^{-1}$.  While enhancement factors vary strongly among the available Skyrme and Gogny functionals \cite{Decharge:1983kx,Meyer:1982fk,Chabanat:1997zt,Bender:2003bs}, our less phenomenological approach consistently favors a limited range of values close to the enhancement factors extracted from experimental photoabsorption cross-section data, which are 70-75\% for $A\gtrsim 100$ \cite{Lepretre:1981uq}. 

\subsection{\label{sec:res_mono}$0^+$ Channel}

Having established the general effects of the SRG evolution on our QRPA results in the previous section, we now discuss the response of the calcium isotopic chain to the transition operators defined in Sec. \ref{sec:qrpa_trans}. Figure \ref{fig:srg0600_dd645_CaXX_ISVM} shows the isoscalar and isovector monopole response for selected calcium isotopes. Since the results for the three SRG parameters are qualitatively similar, we display only the response for $\Vsrg$+DDI with $\lambda=2.02\,\fm^{-1}$ and $C_{3N}=3.87\,\GeV\fm^{6}$ (cf. Tab. \ref{tab:lambda_C3N}). 

\begin{figure}[t]
  \includegraphics[width=\columnwidth]{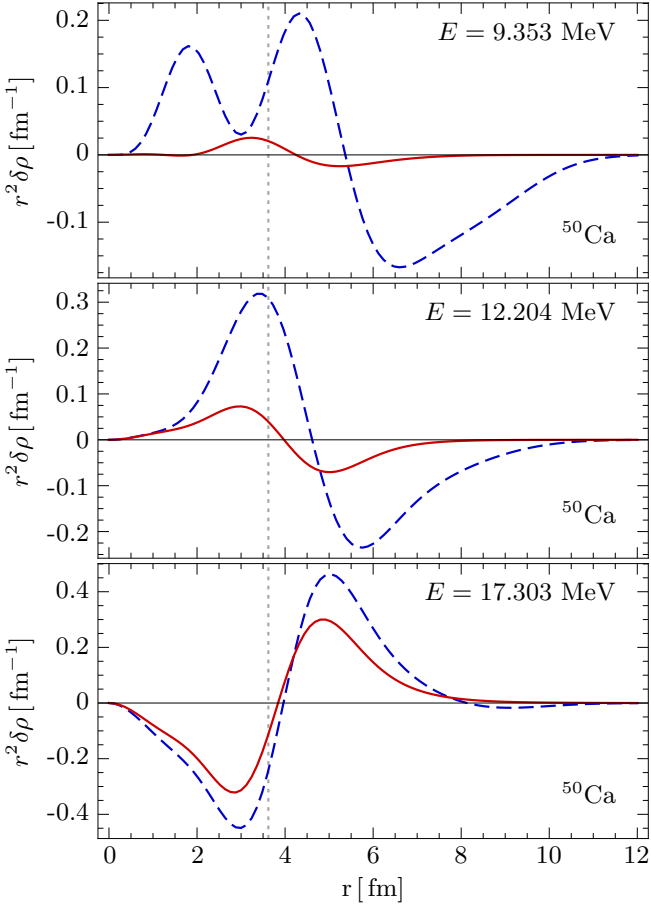}
  \caption{\label{fig:srg0600_dd645_Ca50_ISM_drho}(Color online) 
  Proton (\linemediumsolid[FGRed]) and neutron (\linemediumdashed[FGBlue]) transition densities for the major monopole peaks in $\nuc{Ca}{50}$. The light gray dashed lines indicate the calculated point-nucleon root-mean-square radius. 
  }
\end{figure}

The response exhibits the well-defined isoscalar giant resonance, which is found in the energy interval from $10$ to $20\,\MeV$ in the light calcium isotopes, and shifted to slightly lower energies as $N$ increases. In the isovector channel, we observe a distinct giant resonance peak around $32\,\MeV$ in $\nuc{Ca}{40}$, which fragments into a broad distribution with growing neutron excess. Beyond the major shell closure in $\nuc{Ca}{40}$, pronounced peaks start to emerge in a region around $10\,\MeV$. For the lowest peak, in particular, the isoscalar and isovector response match closely in all isotopes up to $\nuc{Ca}{60}$, indicating that the response is strongly dominated by neutron transitions. As an illustrative example, we show the proton and neutron transition densities of the three major isoscalar peaks in $\nuc{Ca}{50}$ in Fig. \ref{fig:srg0600_dd645_Ca50_ISM_drho}. For the state at $E=9.352\,\MeV$, the proton transition density $\delta\rho_{p}$ is negligibly small compared to the neutron transition density $\delta\rho_{n}$, and $\delta\rho_{n}$ extends far into the nuclear exterior. The state at $E=12.204\,\MeV$ is of transitory character between the neutron-dominated low-lying state and the almost completely isoscalar Giant Monopole Resonance state at $E=17.303$, where $\delta\rho_{p}$ and $\delta\rho_{n}$ have similar extensions. 

While the low-lying strength emerges naturally and becomes enhanced as we move toward the neutron drip line, the existence of low-lying strength in $\nuc{Ca}{48}$ in our calculations is a result of the insufficient spin-orbit splitting of the neutron $0f$ levels, which creates a major shell for $20<N\leq40$ and turns $\nuc{Ca}{48}$ into an open-shell nucleus (cf. Sec. \ref{sec:hfb}). The otherwise qualitatively similar results of the recent Skyrme QRPA survey by Terasaki and Engel \cite{Terasaki:2006fk} do not exhibit a low-lying peak in $\nuc{Ca}{48}$ since the EDFs used in their approach produce the correct major shell closures at $N=20$ and $28$. Furthermore, we observe an overall enhancement of the low-lying excitations compared to Skyrme QRPA, likely due to differences in the single-particle spectra. In the Skyrme QRPA calculations, roughly 10-15\% of the the isoscalar monopole EWSR are exhausted by states below $10\,\MeV$ for isotopes with $N\geq30$ (cf. Fig. 5 of Ref. \cite{Terasaki:2006fk}), while we obtain a growing exhaustion between 10\% in $\nuc{Ca}{50}$ and 20\% in $\nuc{Ca}{60}$ in our calculations. Keep in mind, however, that the details of the running energy-weighted sum depend on the SRG parameter $\lambda$, as discussed in Sec. \ref{sec:res_srg}.
 
\begin{figure}[t]
  \includegraphics[width=\columnwidth]{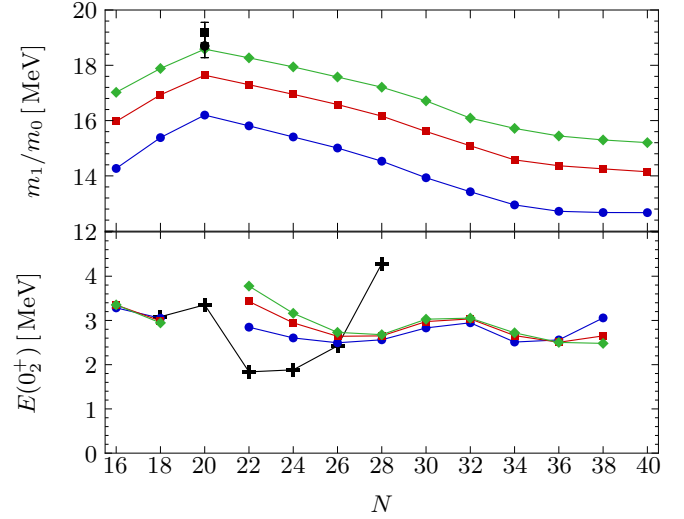}
  \caption{\label{fig:srgXXXX_ddXXX_CaXX_E0}(Color online)
    ISM centroids (top) and energies of the first excited $0^{+}$ states (bottom) in the calcium isotopic chain, for $\Vsrg$+DDI with $(\lambda [\fm^{-1}], C_{3N} [\GeV\fm^6])$= (2.40, 2.94) (\symbolcircle[FGBlue]), (2.02, 3.87) (\symbolbox[FGRed]), and (1.78, 4.41) (\symboldiamond[FGGreen]). Moments of the strength distribution were calculated in the energy interval $5-40\,\MeV$. Experimental data are indicated by black symbols: Centroids for the $\nuc{Ca}{40}$ giant monopole resonance were taken from the two analyses in Ref. \cite{Youngblood:2001hc}, 0+ energies from Ref. \cite{web:NuDat}.
  }
\end{figure}

The upper panel of Fig. \ref{fig:srgXXXX_ddXXX_CaXX_E0} shows the centroids of the ISM response for the three different sets of $\lambda$'s and $C_{3N}$'s. The behavior of the centroid energies as a function of the neutron number is identical for all three $\Vsrg$+DDI, only the energies themselves increase as $\lambda$ is lowered, as discussed in Sec. \ref{sec:res_srg}. The centroids clearly reflect the shift of the ISM strength to lower energies as the neutron excess grows. The trend from $\nuc{Ca}{36}$ to $\nuc{Ca}{40}$ indicates a similar shift in proton-rich calcium isotopes.

\begin{figure*}[t]
  \includegraphics[width=2\columnwidth]{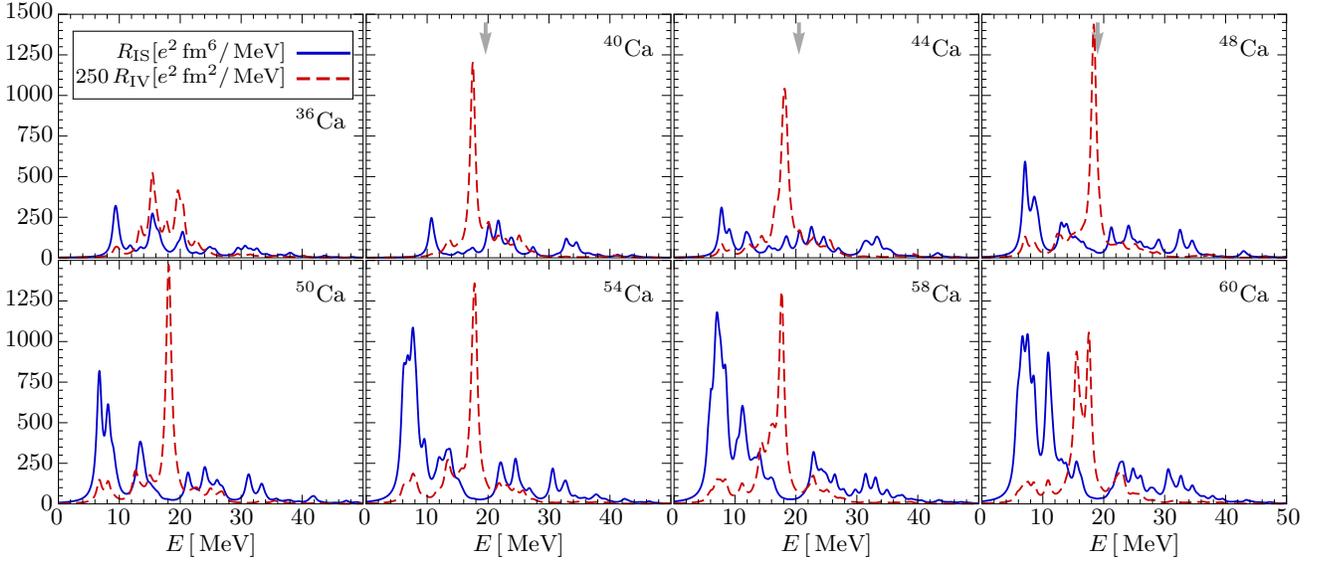}
  \caption{\label{fig:srg0600_dd645_CaXX_ISVD}(Color online)
    Isoscalar (\linemediumsolid[FGBlue])\, and isovector dipole response (\linemediumdashed[FGRed])\, of selected calcium isotopes for $\Vsrg$+DDI with $(\lambda [\fm^{-1}], C_{3N} [\GeV\fm^6])$= (2.02, 3.87). The discrete strength distributions have been folded with a Lorentzian of width $\Gamma=1\,\MeV$, and the isovector strength distribution has been scaled by a factor 250 for visibility. The arrows indicate the main giant dipole resonance peaks from photo-absorption cross-section data \cite{Erokhova:2003dz,web:CDFE}.
  }
\end{figure*}

In the lower panel of Fig. \ref{fig:srgXXXX_ddXXX_CaXX_E0} we display the energies of the first excited $0^{+}$ states. We use the criterion introduced by Terasaki et al. in Ref. \cite{Terasaki:2008zr} and discard states for which
\begin{equation}\label{eq:levcrit}
  \Delta A = 2 \sum_{ij}\left(X_{ij}^{2}-Y_{ij}^{2}\right)\left(u_{i}^{2}-v_{j}^{2}\right)\gtrsim 2\,,
\end{equation}
because these states have pair-transfer character, and carry negligible strength from particle-hole excitations due to the consistency of our QRPA framework. In the closed-shell nuclei $\nuc{Ca}{40}$ and $\nuc{Ca}{60}$ (which has a closed-shell structure in our calculations), we find no low-lying $0^{+}$ state. In the case of $\nuc{Ca}{40}$, experimental data suggest that this state is intrinsically deformed \cite{Ideguchi:2001kx} and can therefore not be described by the present spherical QRPA approach. 

The degree of $\lambda$-dependence of the $0^{+}$ states varies with the considered isotopes, and appears to be connected to the collectivity of each state. For the isotopes with  $26 \leq N \leq 36$, the collectivity of the $0^{+}$ states is enhanced by pairing correlations, and their energies depend only weakly on $\lambda$ in the studied parameter range. The weak variation of the calculated $0^{+}$ energies with the neutron number is qualitatively compatible with the experimental $0^{+}$ energies in the open-shell calcium isotopes between the physical major shell closures at $N=20$ and $N=28$. Thus, we expect to reproduce the experimental trends with a realistic interaction which yields the correct spin-orbit splittings.  

\begin{figure}[t]
  \includegraphics[width=\columnwidth]{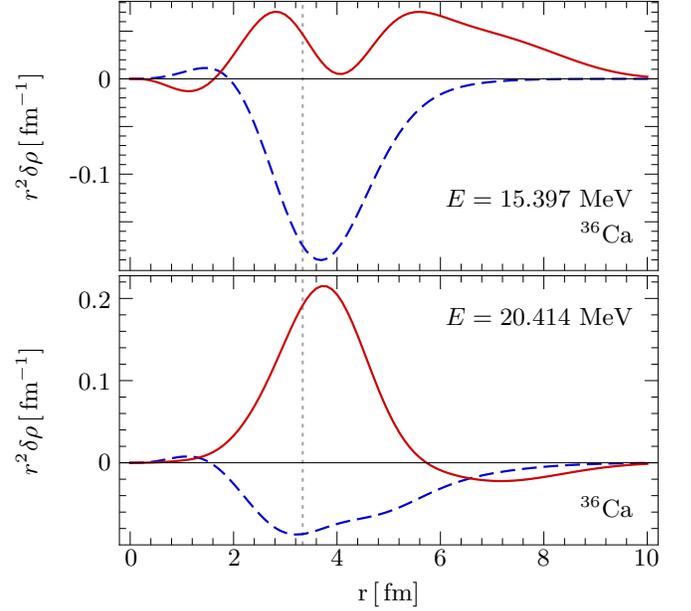}
  \caption{\label{fig:srg0600_dd645_CaXX_drho}(Color online) 
  Proton (\linemediumsolid[FGRed]) and neutron (\linemediumdashed[FGBlue]) dipole transition densities for selected states in $\nuc{Ca}{36}$ (see text). The light gray dashed lines indicate the calculated point-nucleon root-mean-square radius. 
  }
\end{figure}

\subsection{$1^-$ Channel}
In Fig. \ref{fig:srg0600_dd645_CaXX_ISVD}, we show the isoscalar and isovector dipole response of selected calcium isotopes for $\Vsrg$+DDI with $\lambda=2.02\,\fm^{-1}$ and $C_{3N}=3.87\,\GeV\fm^{6}$. The isoscalar response has two major components that correspond to $1\hbar\omega$ and $3\hbar\omega$ excitations, respectively. For $N\geq 20$, these components are well separated energetically by the isovector giant dipole resonance (IVGDR), i.e., states lying between 15 and 20 $\MeV$ carry significant isovector dipole strength, but only little isoscalar dipole strength. In $\nuc{Ca}{36}$, however, the two states at $15.397$ and $20.414\,\MeV$ carry both. Their proton and neutron transition densities are displayed in Fig. \ref{fig:srg0600_dd645_CaXX_drho}. We immediately note that the roles of protons and neutrons are roughly reversed in the two states, while the densities of the states are structurally similar: One density exhibits a pronounced peak near the surface, while the other is wide and flat. Due to the mismatched shapes of $\delta\rho_{n}$ and $\delta\rho_{p}$ in the surface region, both states contribute significantly to the isoscalar and isovector response. 

Returning to Fig. \ref{fig:srg0600_dd645_CaXX_ISVD}, we note that the isoscalar response is further split into two components in the two previously identified regions. A low-lying peak around $10\,\MeV$ is almost completely produced by proton transitions in $\nuc{Ca}{36}$ and $\nuc{Ca}{38}$ (not shown). In $\nuc{Ca}{40}$, neutron transitions start to contribute about $30\%$ of the state's norm, making this mode largely isoscalar \cite{Papakonstantinou:2011cr}, while beyond the $N=20$ shell closure, neutrons completely dominate this lower portion of the low-lying strength. The second peak at $\sim13\,\MeV$ that emerges beyond $N=20$ also contains proton contributions, starting around $30-35\%$ in $\nuc{Ca}{42}$ and decreasing below 10\% in $\nuc{Ca}{60}$. The peaks grow more pronounced and eventually overlap with growing neutron excess, and the centroid of their strength is shifted to lower energies. 

In the high-lying region from $20$ to $35\,\MeV$, there is a very broad structure corresponding to the isoscalar giant dipole resonance (ISGDR). While Fig. \ref{fig:srg0600_dd645_CaXX_ISVD} suggests that there are two distinct groups of peaks, the states of both groups contain balanced contributions from proton and neutron transitions, as expected for the ISGDR, and there are no indications of major structural differences.

\begin{figure}[t]
  \includegraphics[width=\columnwidth]{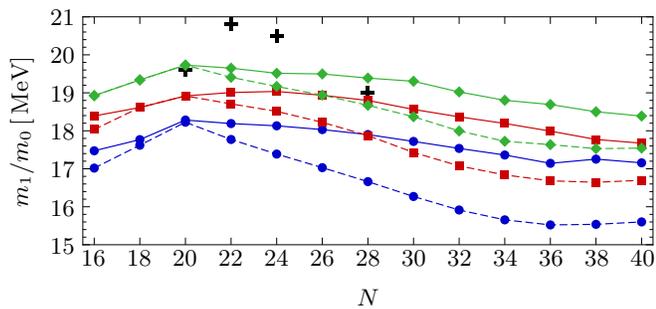}
  \caption{\label{fig:srgXXXX_ddXXX_CaXX_IVD_cent}(Color online)
    Centroids of the IVD response in calcium isotopes for $\Vsrg$+DDI with $(\lambda [\fm^{-1}], C_{3N} [\GeV\fm^6])$= (2.40, 2.94) (\symbolcircle[FGBlue]), (2.02, 3.87) (\symbolbox[FGRed]), and (1.78, 4.41) (\symboldiamond[FGGreen]). Moments $m_{i}$ were calculated in the energy intervals $10-40\,\MeV$ (\linemediumsolid)\, and $0-40\,\MeV$(\linemediumdashed), respectively. Experimental GDR peaks from photoabsorption data \cite{Erokhova:2003dz,web:CDFE} are indicated by (\symbolcross). 
  }
\end{figure}

As already mentioned above, the isovector dipole response of all calcium isotopes exhibits the pronounced giant dipole resonance (GDR) between $15$ and $20\,\MeV$, and the energies of the main resonance peaks agree reasonably well with experimental values extracted from photoabsorption data \cite{web:CDFE}. The corresponding GDR centroid energies, which are shown in Fig. \ref{fig:srgXXXX_ddXXX_CaXX_IVD_cent}, vary only weakly with $N$. The centroids exhibit the $\lambda$-dependence we expect from our discussion in Sec. \ref{sec:res_srg}, i.e., the centroid energy increases as $\lambda$ is lowered from $2.40$ to $1.78\,\fm^{-1}$. 

With increasing neutron excess, the IVD response develops a pronounced low-lying peak structure below $10\,\MeV$. A similar but less pronounced peak is observed in the proton-rich nuclei $\nuc{Ca}{36}$ (cf. Fig. \ref{fig:srg0600_dd645_CaXX_ISVD}) and $\nuc{Ca}{38}$ (not shown). The emergence of low-lying $E1$ strength, or pygmy dipole strength as it is usually called, is highlighted in Fig. \ref{fig:srgXXXX_ddXXX_CaXX_IVD_cent} by including the centroid energy for the total IVD strength up to $40\,\MeV$ rather than just the region of the GDR. It is a robust feature under variations of the SRG parameter $\lambda$, but the details may differ because strength is shifted beyond $10\,\MeV$ as $\lambda$ decreases.

For the low-lying E1 strength of $\nuc{Ca}{40}$, $\nuc{Ca}{44}$, and $\nuc{Ca}{48}$, experimental data are available from measurements conducted at the S-DALINAC \cite{Hartmann:2004fu}. In general our calculations overestimate the energy of the PDRs and tend to underestimate the strength below 10 MeV. The overestimation of the states' energy is ultimately in line with the conclusions from other studies in the literature, namely that in order to properly describe the low-lying E1 strength one needs to go beyond (Q)RPA and include particle-phonon coupling, as in, e.g., Ref. \cite{Hartmann:2004fu}.

\begin{figure}[t]
  \includegraphics[width=\columnwidth]{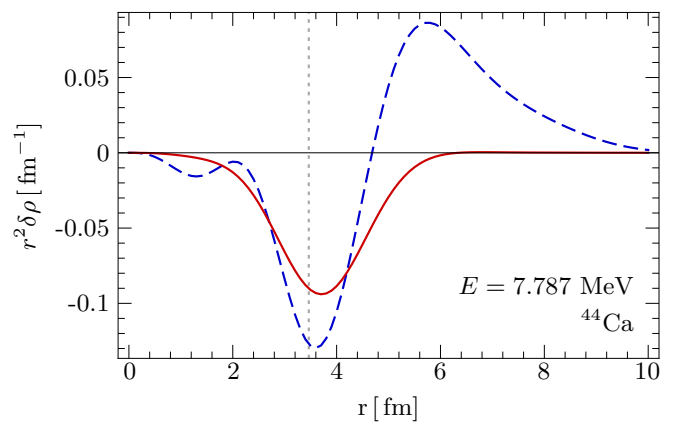}
  \caption{\label{fig:srg0600_dd645_Ca44_drho}(Color online)
  Proton (\linemediumsolid[FGRed]) and neutron (\linemediumdashed[FGBlue]) dipole transition densities for the low-lying state at $E=7.787\,\MeV$ in $\nuc{Ca}{44}$. The light gray dashed line indicates the calculated point-nucleon root-mean square radius. 
  }
\end{figure}

From $\nuc{Ca}{42}$ to $\nuc{Ca}{60}$, states below $10\,\MeV$ carry between 2\% and 4\% of the E1 EWSR, corresponding to 4\%-7\% of the TRK sum rule. The percentage of the sum rule increases smoothly with growing neutron excess, as expected at the QRPA level \cite{Hartmann:2004fu}. Almost all of the low-lying states carry both isoscalar and isovector dipole strength. As indicated in the discussion above, 70-80\% of the norm of the low-lying states in $\nuc{Ca}{40}$ is made up of proton transitions, while the low-lying states of the nuclei with $N>20$ are completely dominated by neutrons, which contribute 95-99\% of the individual states' norms. Further analysis also shows that typically 10-20 different transitions contribute 1\% or more each to the norm of each low-lying state, which confirms their collective nature, and supports their interpretation as pygmy dipole modes. The proton and neutron transition densities of the strongest low-lying state in $\nuc{Ca}{44}$ that are shown in Fig. \ref{fig:srg0600_dd645_Ca44_drho} exhibit the pygmy mode characteristics: $\delta\rho_{p}$ and $\delta\rho_{n}$ are in phase at short ranges, and out of phase just beyond the surface region of the nucleus, where $\delta\rho_{n}$ also has a pronounced neutron tail, in agreement with other studies \cite{Paar:2007ss,Tsoneva:2008il,Co:2009vn}.

\subsection{$2^+$ Channel}
\begin{figure*}[t]
  \includegraphics[width=2\columnwidth]{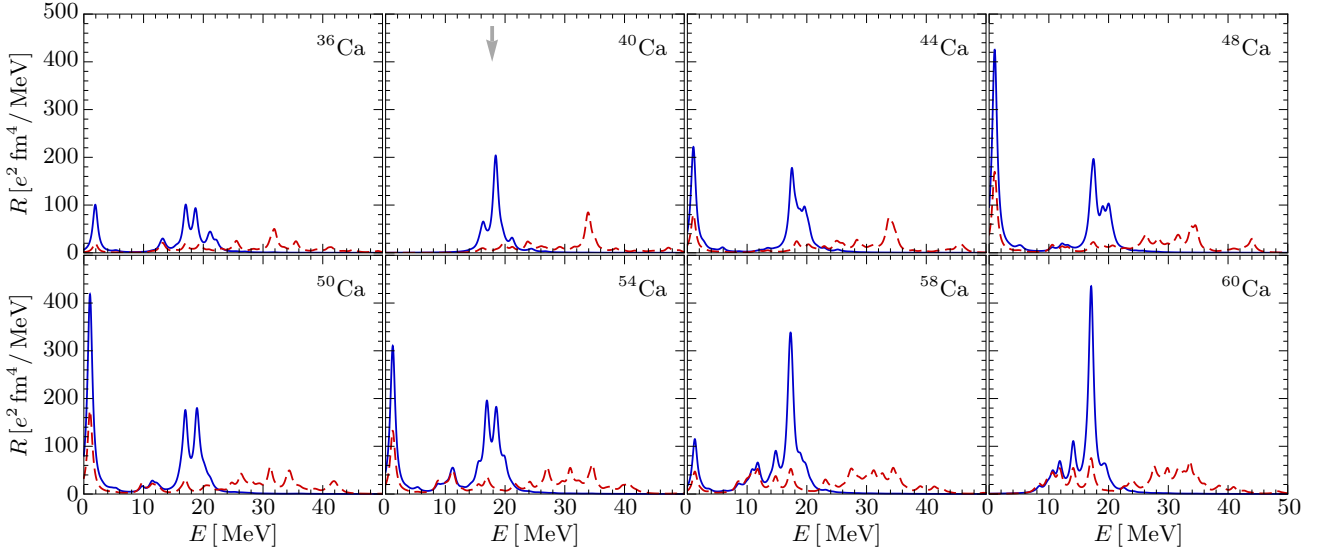}
  \caption{\label{fig:srg0600_dd645_CaXX_ISVQ}(Color online)
    Isoscalar (\linemediumsolid[FGBlue]) and isovector quadrupole response (\linemediumdashed[FGRed]) of selected calcium isotopes for $\Vsrg$+DDI with $(\lambda [\fm^{-1}], C_{3N} [\GeV\fm^6])$= (2.02, 3.87). The discrete strength distributions have been folded with a Lorentzian of width $\Gamma=1\,\MeV$. The arrow indicates the experimental centroid from Ref. \cite{Youngblood:2001hc}.
  }
\end{figure*}

In Fig. \ref{fig:srg0600_dd645_CaXX_ISVQ}, we show the isoscalar and isovector quadrupole response of selected calcium isotopes for $\Vsrg$+DDI with $\lambda=2.02\,\fm^{-1}$ and $C_{3N}=3.87\,\GeV\fm^{6}$. We find the pronounced isoscalar giant quadrupole resonance (ISGQR) between 15 and 20$\,\MeV$, which exhibits moderate fragmentation. The isovector response is very broad in comparison, with the bulk of the strength residing in the IVGQR above $25\,\MeV$. 

\begin{figure}[t]
  \includegraphics[width=\columnwidth]{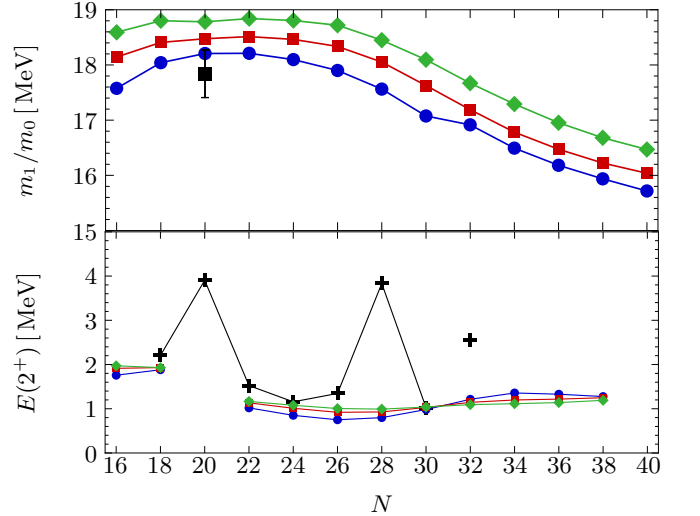}
  \caption{\label{fig:srgXXXX_ddXXX_CaXX_E2}(Color online)
    ISQ centroids (top), and energies of the first $2^{+}$ states (bottom) in the calcium isotopic chain, for $\Vsrg$+DDI with $(\lambda [\fm^{-1}], C_{3N} [\GeV\fm^6])$= (2.40, 2.94) (\symbolcircle[FGBlue]), (2.02, 3.87) (\symbolbox[FGRed]), and (1.78, 4.41) (\symboldiamond[FGGreen]). Moments of the strength distribution were calculated in the energy interval $8-30\,\MeV$. Experimental data are indicated by black symbols: the centroid for the $\nuc{Ca}{40}$ giant quadrupole resonance was taken from Ref. \cite{Youngblood:2001hc} and 2+ energies from Ref. \cite{web:NuDat}. The $2^{+}$ assignment is still tentative for the level in $\nuc{Ca}{52}$.
  }
\end{figure}

In the upper panel of Fig. \ref{fig:srgXXXX_ddXXX_CaXX_E2}, we show the centroid energies of the ISGQR for the three studied SRG parameters. While the centroids exhibit the same trends under variation of $\lambda$ as the ISM and IVD centroids, we note that the resulting change in energies is smaller, covering an interval of roughly $1\,\MeV$ compared to the $\sim2\,\MeV$ range in the other cases. This observation matches the reduced sensitivity of the ISQ EWSR to variations of $\lambda$ discussed in Sec. \ref{sec:res_srg}. We also note that contrary to the ISGMR and IVGDR cases, where we typically approach the experimental centroid or peak energies from below as $\lambda$ is lowered, we overestimate the experimental centroid energy of the ISGQR in $\nuc{Ca}{40}$ already for $\lambda=2.40\,\fm^{-1}$. 

Figure \ref{fig:srg0600_dd645_CaXX_ISVQ} exhibits prominent peaks at low energies, corresponding to the lowest $2^{+}$ states of the respective nuclei. In the lower panel of Fig. \ref{fig:srgXXXX_ddXXX_CaXX_E2} we show the energies of these states, which have been distinguished from states with pair-transfer character using the criterion \eqref{eq:levcrit} (cf. Sec. \ref{sec:res_mono}). A recent survey using a spherical Skyrme QRPA approach can be found in \cite{Terasaki:2008zr}. Similar to the first excited $0^{+}$ states, the lowest $2^{+}$ states are insensitive to variations in the SRG parameter $\lambda$, but for the latter, this also holds in the isotopes next to the major shell closures (cf. Fig. \ref{fig:srgXXXX_ddXXX_CaXX_E0}). In $\nuc{Ca}{40}$ and $\nuc{Ca}{60}$, we do not find $2^{+}$ states below $10\,\MeV$. The absence of the experimentally observed level in $\nuc{Ca}{40}$ suggests that this state has a structure that cannot be described properly by our spherical QRPA. 

We find reasonable agreement between our results and the experimental $2^{+}$ energies in the open-shell calcium isotopes. In semimagic spherical open-shell nuclei, the $2^{+}$ states lie at nearly constant energies along isotopic chains due to pairing correlations \cite{Talmi:2003kx}. In our calculations, all isotopes from $\nuc{Ca}{42}$ to $\nuc{Ca}{58}$ exhibit this characteristic behavior because the insufficient spin-orbit splitting of the $0f$ levels creates a major shell with $20<N\leq40$. Consequently, we do not reproduce the increased experimental $2^{+}$ energies in $\nuc{Ca}{40}$ and $\nuc{Ca}{48}$, which signal the natural major shell closures, nor the experimentally observed level at $E=2.563\,\MeV$ in $\nuc{Ca}{52}$, which has been tentatively identified as a $2^{+}$ state and whose increased energy is interpreted as a sign of an enhanced $1p_{3/2}$ sub-shell closure \cite{Huck:1985uq}.

We conclude our discussion of the quadrupole response by focusing on the peak near $10\,\MeV$, which emerges in the quadrupole strength distributions of neutron-rich calcium isotopes (cf. Fig. \ref{fig:srg0600_dd645_CaXX_ISVQ}). 
All of the discrete states between $8$ and $12\,\MeV$ contributing to this peak carry isoscalar and isovector strength in almost equal amounts because they are dominated by neutron transitions. They are also collective, receiving significant contributions from about 10 different transitions each, suggesting that they can be interpreted as pygmy quadrupole modes, whose existence was recently proposed in Ref. \cite{Tsoneva:2011ys}.

\begin{figure}[t]
  \includegraphics[width=\columnwidth]{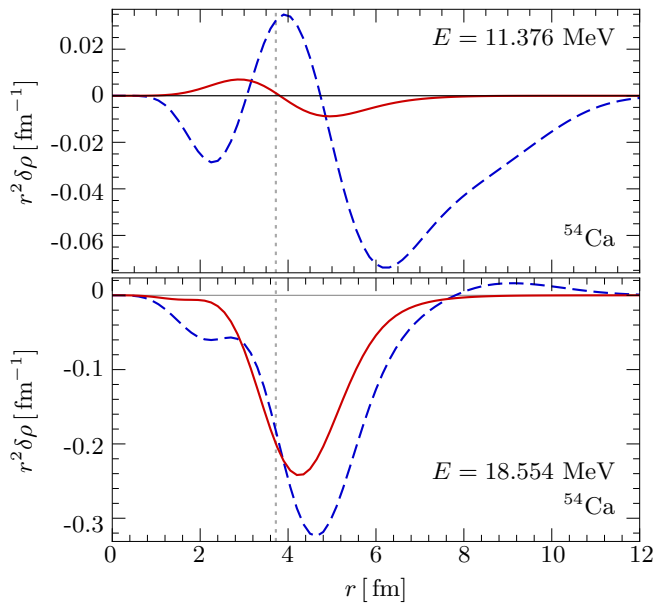}
  \caption{\label{fig:srg0600_dd645_Ca54_drho}(Color online) 
  Proton (\linemediumsolid[FGRed]) and neutron (\linemediumdashed[FGBlue]) quadrupole transition densities for a Pygmy (top) and a GQR mode (bottom) in $\nuc{Ca}{54}$. The light gray dashed lines indicate the calculated point-nucleon root-mean square radius. 
  }
\end{figure}

In Fig. \ref{fig:srg0600_dd645_Ca54_drho}, we compare the transition densities of the strongest $2^{+}$ state from the peak region to those of the main ISGQR state in $\nuc{Ca}{54}$. The transition densities of the ISGQR are in phase up to $\sim7.5\,\fm$, and have a similar magnitude. Their extrema are located within $1\,\fm$ of the calculated point-nucleon radius of $\nuc{Ca}{54}$. The proton transition density of the pygmy mode has a very small amplitude but similar extension to that of the ISGQR. The neutron transition density, on the other hand, extends much further, and the bulk of its contribution to the transition operator is generated at much larger distances than for the ISGQR state, which supports its interpretation as a neutron-skin excitation. 

\section{Conclusions}
We have developed a fully consistent QRPA framework for arbitrary $NN$ interactions, represented by their two-body matrix elements, as well as $3N$ contact interactions via equivalent linearly density-dependent interactions. We use an intrinsic Hamiltonian  and treat the Coulomb interaction exactly. The QRPA is built on the HFB ground states obtained with the code described in Ref. \cite{Hergert:2009zn}. Since we do not truncate the QRPA configuration space and use the same Hamiltonian in the HFB and QRPA calculations, we achieve an excellent decoupling of the spurious strength. 

In the present work, we have employed our QRPA framework to study the nuclear response using $NN$ interactions derived from Argonne V18 by means of the Similarity Renormalization Group, supplemented by a density-dependent interaction whose strength $C_{3N}$ is fit to the charge radii of closed-shell nuclei. By refitting $C_{3N}$ for each value of the SRG parameter $\lambda$, we have been able to absorb the effect of the SRG evolution on the charge radii at the HFB level. 

Although the HFB ground-state energy strongly depends on $\lambda$, and binding energy on the order of $4-6\,\MeV$ per nucleon is missing, the theoretical single-particle spectra around the Fermi surface are reasonable, aside from the underestimation of the spin-orbit splittings, which indicates the need for additional spin-orbit strength from the $3N$ interaction. Experimental odd-even binding energy differences, the analogs of pairing gaps in finite nuclei, are reasonably close to experimental values. Likewise, we obtain reasonable properties for the collective dynamics of the studied nuclei. 

In the nuclear response, we have identified two basic effects of the SRG evolution: The response is shifted to slightly higher energies as $\lambda$ is lowered, plausibly because interaction strength is shifted to momentum-dependent terms in the interaction, resulting in a reduction of the effective mass, and an increased spreading of the major shells in the single-particle spectrum. At the same time, the levels within a major shell are bunched closer together, resulting in less fragmentation at the (Q)RPA level. The important energy-weighted sum rules are satisfied for the studied SRG interactions, and exhibit only weak dependence on the SRG parameter $\lambda$. Interestingly, we found that the enhancement of the isovector dipole EWSR over the classical Thomas-Reiche-Kuhn value seems to be consistently close to experimental data, whereas the theoretical values strongly depend on the used EDF in Skyrme of Gogny (Q)RPA studies.

Our results for the monopole, dipole, and quadrupole response of the calcium isotopes are comparable to existing studies based on phenomenological EDFs. We have obtained reasonable centroid energies for the important giant resonances and identified multiple low-lying states in the dipole and quadrupole response that exhibit the characteristics of Pygmy modes. In accordance with other studies in the literature, the energies of these states are overestimated at the (Q)RPA level, illustrating the need to include many-body effects like quasiparticle-phonon coupling to achieve a proper description of low-lying strength. 

The next major development stage of our QRPA framework will be the switch to SRG-evolved chiral $NN$ and $3N$ interactions. Since the latter contain spin-orbit and tensor structures, we expect an impact on the issue of the underestimated spin-orbit splittings, although it remains to be seen whether the problem will be fixed. Realistic spin-orbit splittings are essential to allow studies of heavier isotopic chains, where the ground states develop a deformation for the currently used $\Vsrg$+DDI, causing the breakdown of the spherical QRPA.

The extension of our HFB+QRPA framework to deformed nuclei is an obvious direction for further research, but the treatment of realistic $NN$ and $3N$ interactions in deformed bases will require significant work, particularly on the transformation between relative partial waves and uncoupled deformed single-particle states. Besides the treatment of deformation, we are looking into extensions toward more complicated configurations like $4qp$ excitations, i.e., a Second QRPA analogous to the Second RPA described in Ref. \cite{Papakonstantinou:2010oq}, and quasiparticle-phonon coupling.

Our initial QRPA survey of the calcium isotopes has indicated interesting structural features like low-lying modes, including the prominent $1\hbar\omega$ peak in the isoscalar dipole response, which was studied for closed-shell nuclei in Ref. \cite{Papakonstantinou:2011cr}. More detailed studies of the physics of these modes in open-shell nuclei will be the subject of research in the near future.

\section*{Acknowledgments}
We thank J. Terasaki for providing us with his spherical QRPA code for comparison purposes and useful comments.

H. H. is supported by the National Science Foundation under Grant No. PHY-0758125, and the UNEDF SciDac Collaboration under the U.S. Department of Energy Grant No. DE-FC02-09ER41585.
P. P. and R. R. are supported by the Deutsche Forschungsgemeinschaft through contract SFB 634, by the Helmholtz International Center for FAIR within the framework of the LOEWE program launched by the State of Hesse, and the German Federal Ministry of Education and Research (BMBF 06DA9040I).

\appendix
\section{Explicit Form of the QRPA Matrices \label{app:qrpa_mat}}
The matrices $A$ and $B$ are given by [$\mu=(n_\mu l_\mu j_\mu\tau_\mu)$]
\begin{align}
  &A^J_{\mu\mu',\nu\nu'}
      =\frac{1}{\sqrt{1+\delta_{\mu\mu'}}}\frac{1}{\sqrt{1+\delta_{\nu\nu'}}}
	\Big\{
	      H^{11}_{\mu\nu}\delta_{\mu'\nu'}+H^{11}_{\mu'\nu'}\delta_{\mu\nu}\notag\\[3pt]
        &\quad-(-1)^{j_\mu+j_{\mu'}-J}
	     \left(H^{11}_{\mu'\nu}\delta_{\mu\nu'}+
                   H^{11}_{\mu\nu'}\delta_{\mu'\nu}\right)\notag\\[3pt]
        &\quad
             +F(\mu\mu'\nu\nu';J)
	     \left(u_\mu v_{\mu'}u_\nu v_{\nu'}+ (u \leftrightarrow v)\right)\notag\\[3pt]
        &\quad
             -(-1)^{j_\nu+j_{\nu'}-J}F(\mu\mu'\nu'\nu;J)
	     \left(u_\mu v_{\mu'}v_\nu u_{\nu'}+(u \leftrightarrow v)\right)\notag\\[3pt]
         &\quad
	      +G(\mu\mu'\nu\nu';J)
	      \left(u_\mu u_{\mu'}u_\nu u_{\nu'}+ (u \leftrightarrow v)\right)\Big\}
\end{align}
and 
\begin{align}
  &B^J_{\mu\mu',\nu\nu'}=\frac{1}{\sqrt{1+\delta_{\mu\mu'}}}\frac{1}{\sqrt{1+\delta_{\nu\nu'}}}\notag\\
        &\hphantom{=}
        \times\Big\{
             F(\mu\mu'\nu\nu';J)\left(v_\mu u_{\mu'}u_\nu v_{\nu'}+(u \leftrightarrow v) \right)
            \notag\\
         &\hphantom{=\bigg(\;}
             -(-1)^{j_\nu+j_{\nu'}-J}F(\mu\mu'\nu'\nu;J)\left(u_\mu v_{\mu'}u_\nu v_{\nu'}+ (u \leftrightarrow v)\right)\notag\\
         &\hphantom{=\bigg(\;}
             -G(\mu\mu'\nu\nu';J)\left(u_\mu u_{\mu'}v_\nu v_{\nu'}+ (u \leftrightarrow v)\right)\Big\}\,.
\end{align}

The single-quasiparticle term $H^{11}_{\mu\mu'}$ reads
\begin{equation}\label{eq:H11}
  H^{11}_{\mu\nu}=(u_\mu u_\nu - v_\mu v_\nu)(h_{\mu\nu}-\lambda\delta_{\mu\nu}) + (u_\mu v_\nu + v_\mu u_\nu)\Delta_{\mu\nu}\,,
\end{equation}
where $\lambda$ is the chemical potential (separate for neutrons and protons), 
\begin{equation}\label{eq:def_h}
   h_{\mu\mu'} = \left(1-\frac{1}{A}\right)t_{\mu\mu'} + \sum_{\nu}(2j_\nu+1)\bar{v}_{\mu\nu\mu'\nu}v^2_{\nu}
\end{equation}
is the particle-hole field, and
\begin{equation}
  \Delta_{\mu\mu'} = \frac{1}{2}\sum_{\nu}(2j_\nu+1)\bar{v}_{\mu\mu'\nu\nu}u_\nu v_{\nu}
\end{equation}
is the pairing field in the canonical basis representation. Note that both $h$ and $\Delta$ are diagonal in $l$ and $j$ and independent of magnetic quantum numbers due to parity and spherical symmetry, and the two-body matrix elements $\bar{v}_{\mu\mu'\nu\nu'}$ contain all two-body terms of the Hamiltonian, i.e., nuclear and electromagnetic interactions, as well as the two-body center-of-mass correction \cite{Hergert:2009zn,Hergert:2009wh}.

We adopt the notation of Ref. \cite{Terasaki:2005ja} for the particle-hole and particle-particle channel matrix elements, which are given by
\begin{align}
  F(\mu\mu'\nu\nu';J)
  &=\sum_{J'}(-1)^{j_{\mu'}+j_\nu+J'}
    (2J'+1)\sixj{j_\mu & j_{\mu'} & J \\ j_\nu &j_{\nu'} &J'}\notag\\
&\qquad\times\matrixe{\mu\nu';J'}{\vO}{\mu'\nu;J'}\label{eq:def_F}
\end{align}
and
\begin{equation}
  G(\mu\mu'\nu\nu';J)=\matrixe{\mu\mu';J}{\vO}{\nu\nu';J}\,.
\end{equation}
As stated in Sec. \ref{sec:qrpa}, the two-body matrix elements are antisymmetrized but not normalized. 

\section{Density-Dependent Interaction \& Rearrangement Terms \label{app:ddi}}
The contribution of the density-dependent interaction \eqref{eq:def_ddi} to the energy is given by
\begin{align}\label{eq:ddi_energy}
  E_\rho&=\frac{\dmatrixe{\Psi}{\VO[\rho]}}{\braketn{\Psi}}
            =\frac{\dmatrixe{\Psi}{\sum_{i<j}\vO_{ij}[\rho]}}{\braketn{\Psi}}\notag\\
           &=C_{3N} \pi\int dr\, r^2\rho(r)\rho_n(r)\rho_p(r)
\end{align}
where
\begin{equation}
   \rho_\tau(r)=\sum^{\tau=\tau_\mu}_\mu\frac{2j_\mu+1}{4\pi}v_\mu^2 |R_\mu(r)|^2\,,
\end{equation}
$R_\mu(r)$ are the canonical wavefunctions, and the total density is
\begin{equation}
   \rho(r) = \rho_n(r)+\rho_p(r)\,.
\end{equation}
Note that there is no contribution to the pairing energy, since the interaction acts only in the $(S,T)=(1,0)$ channel, i.e., like-particle matrix elements vanish.

Variation of Eq. \eqref{eq:ddi_energy} w.r.t. to the density matrix yields the particle-hole field that has to be added to $h_{\mu\mu'}$ [Eq. \eqref{eq:def_h}]:
\begin{align}
  \Gamma_{\mu\mu'}&=\frac{C_{3N}}{4}\int dr\,r^2R^*_\mu(r)\rho(r)\left[\rho(r)-\rho_{\tau_\mu}(r)\right]R_{\mu'}(r)\notag\\
  &\hphantom{=}+\frac{C_{3N}}{4}\int dr\,r^2R^*_\mu(r)\rho_p(r)\rho_n(r)R_{\mu'}(r)\,,
\end{align}
where the second line is the rearrangement term.

In the QRPA matrices, the DDI contributes the matrix elements
\begin{equation}
   \matrixe{\mu m_{\mu} \nu m_{\nu}}{\vO^{ph}[\rho]}{\mu' m_{\mu'} \nu' m_{\nu'}} = \frac{\partial^2 E_\rho}{\partial \rho_{\mu' m_{\mu'}\mu m_{\mu}} \partial \rho_{\nu' m_{\nu'}\nu m_{\nu}}}
\end{equation}
to the particle-hole interaction \cite{Waroquier:1987ce}, which can be plugged into Eq. \eqref{eq:def_F}. Denoting the normal and rearrangement matrix elements $F_0$ and $F_1$, respectively, one obtains for identical isospins ($\tau\equiv\tau_\mu=\tau_{\mu'}=\tau_\nu=\tau_{\nu'}$)
\begin{equation}
  F_0(\mu\mu'\nu\nu'; J)=0\,,\label{eq:F0_tautau}
\end{equation}
and ($\hat{j}\equiv\sqrt{2j+1}$)
\begin{align}
  &F_1(\mu\mu'\nu\nu')\notag\\
  &= C_{3N}
  \frac{\hjmu\hjjmu\hjnu\hjjnu}{32\pi\hJ^2}\braket{\jmu\half \jjmu-\half}{J0}
   \braket{\jnu\half \jjnu-\half}{J0}\notag\\
&\quad\times   
   (-1)^{\jmu-\jnu}\left(1+(-1)^{\lmu+\llmu+J}\right)\left(1+(-1)^{\lnu+\llnu+J}\right)\notag\\
&\quad\times\int dr\,r^2R_\mu(r)R_{\mu'}(r)R_\nu(r)R_{\nu'}(r)
         \Big[\rho(r)-\rho_\tau(r)\Big]
   \,.\label{eq:F1_tautau}
\end{align}
If the isospins are only pairwise identical ($\tau_\mu=\tau_{\mu'}\neq\tau_\nu=\tau_{\nu'}$),
\begin{align}
   &F_0(\mu\mu'\nu\nu'; J)\notag\\
   &=\frac{C_{3N}}{12}K(\mu\mu'\nu\nu';J)(-1)^{\jmu+\jnu+\jjmu+\jjnu}\notag\\
   &\quad\times\Big\{
      \braket{\jmu\half\jjmu-\half}{J0}\braket{\jnu\half\jjnu-\half}{J0}\notag\\
   &\qquad\qquad\times(-1)^{\jjmu-\jjnu}   
      \left((-1)^{J+\lnu+\llnu}   
       + 2\right)\notag\\
   &\qquad\quad+\braket{\jmu\half\jjmu\half}{J1}\braket{\jnu\half\jjnu\half}{J1}
       (-1)^{\llmu+\llnu}
       \Big\}\label{eq:F0_tautaup}\,,
\end{align}
and
\begin{align}
  &F_1(\mu\mu'\nu\nu')\notag\\
  & = \frac{C_{3N}}{16}
\braket{\jmu\half \jjmu-\half}{J0}
   \braket{\jnu\half \jjnu-\half}{J0}\notag\\
&\quad\times   
   (-1)^{\jmu-\jnu}\left(1+(-1)^{\lmu+\llmu+J}\right)\left(1+(-1)^{\lnu+\llnu+J}\right)\notag\\
&\quad\times
   K(\mu\mu'\nu\nu';J)\,,\label{eq:F1_tautaup}
\end{align}
where the radial integral is given by
\begin{align}
 &K(\mu\mu'\nu\nu';J)\notag\\
  &=\frac{\hjmu\hjjmu\hjnu\hjjnu}{4\pi\hJ^2}\int dr\,r^2R_\mu(r)R_{\mu'}(r)R_\nu(r)R_{\nu'}(r)\rho(r)\,.
\end{align}
We have confirmed that these matrix elements match the more general versions derived in \cite{Waroquier:1987ce} analytically and numerically.


%

\end{document}